\documentclass[sigconf,screen]{acmart}
\usepackage{balance}

\usepackage{subfigure}
\usepackage{multirow}
\usepackage{colortbl}
\usepackage{xspace}
\usepackage{bm}
\usepackage{tcolorbox}
\usepackage{enumitem}
\newcommand{\zz}[1]{#1}

\newcommand{\ourtool}{SARIF\xspace}
\newcommand{\tool}{\ourtool}
\newcommand{\mj}[1]{#1}

\usepackage{caption}
\newcommand{\distance}{7pt}
\DeclareRobustCommand{\mybox}[2][gray!20]{%
\begin{tcolorbox}[
        left=0pt,
        right=0pt,
        top=0pt,
        bottom=0pt,
        colback=#1,
        colframe=#1,
        width=\linewidth, 
        enlarge left by=0mm,
        boxsep=5pt,
        arc=0pt,outer arc=0pt,
        ]
        #2
\end{tcolorbox}
}

\newcommand{\aaj}{$a2a_{adj}$\xspace}
\newcommand{\ccg}{$c2c_{cvg}$\xspace}
\AtBeginDocument{%
  \providecommand\BibTeX{{%
    \normalfont B\kern-0.5em{\scshape i\kern-0.25em b}\kern-0.8em\TeX}}}

\setcopyright{acmlicensed}
\acmPrice{}
\acmDOI{10.1145/3611643.3616285}
\acmYear{2023}
\copyrightyear{2023}
\acmSubmissionID{fse23main-p372-p}
\acmISBN{979-8-4007-0327-0/23/12}
\acmConference[ESEC/FSE '23]{Proceedings of the 31st ACM Joint European Software Engineering Conference and Symposium on the Foundations of Software Engineering}{December 3--9, 2023}{San Francisco, CA, USA}
\acmBooktitle{Proceedings of the 31st ACM Joint European Software Engineering Conference and Symposium on the Foundations of Software Engineering (ESEC/FSE '23), December 3--9, 2023, San Francisco, CA, USA}
\received{2023-02-02}
\received[accepted]{2023-07-27}

\ccsdesc[500]{Software and its engineering~Software maintenance tools}
\ccsdesc[500]{Software and its engineering~Software reverse engineering}
\ccsdesc[500]{Software and its engineering~Maintaining software}

\keywords{software architecture recovery, software module clustering,  reverse engineering, architecture comparison}

\begin{document}

\title{Software Architecture Recovery with Information Fusion}

%
\author{Yiran Zhang}
\affiliation{%
  \institution{Nanyang Technological University}
  \country{Singapore}
}
\email{yiran002@e.ntu.edu.sg}

\author{Zhengzi Xu}
\affiliation{%
  \institution{Nanyang Technological University}
  \country{Singapore}
}
\email{zhengzi.xu@ntu.edu.sg}

\author{Chengwei Liu}
\authornote{Chengwei Liu is the corresponding author}
\affiliation{%
  \institution{Nanyang Technological University}
  \country{Singapore}
}
\email{chengwei001@e.ntu.edu.sg}

\author{Hongxu Chen}
\affiliation{%
  \institution{Huawei Technologies Co., Ltd}
  \city{Shenzhen}
  \country{China}
}
\email{chenhongxu5@huawei.com}

\author{Jianwen Sun}
\affiliation{%
  \institution{Huawei Technologies Co., Ltd}
  \city{Shenzhen}
  \country{China}
}
\email{sunjianwen4@huawei.com}

\author{Dong Qiu}
\affiliation{%
  \institution{Huawei Technologies Co., Ltd}
  \city{Shenzhen}
  \country{China}
}
\email{dong.qiu@huawei.com}

\author{Yang Liu}
\affiliation{%
  \institution{Nanyang Technological University}
  \country{Singapore}
}
\email{yangliu@ntu.edu.sg}

\begin{abstract}
Understanding the architecture is vital for effectively maintaining and managing large software systems. However, as software systems evolve over time, their architectures inevitably change. To keep up with the change, architects need to track the implementation-level changes and update the architectural documentation accordingly, which is time-consuming and error-prone. Therefore, many automatic architecture recovery techniques have been proposed to ease this process. 
Despite efforts have been made to improve the accuracy of architecture recovery, existing solutions still suffer from two limitations. First, most of them only use one or two type of information for the recovery, ignoring the potential usefulness of other sources. Second, they tend to use the information in a coarse-grained manner, overlooking important details within it.

To address these limitations, we propose SARIF, a fully automated architecture recovery technique, which incorporates three types of comprehensive information, including dependencies, code text and folder structure. SARIF can recover architecture more accurately by thoroughly analyzing the details of each type of information and adaptively fusing them based on their relevance and quality. 
To evaluate SARIF, we collected six projects with published ground-truth architectures and three open-source projects labeled by our industrial collaborators. 
We compared SARIF with nine state-of-the-art techniques using three commonly-used architecture similarity metrics and two new metrics. The experimental results show that SARIF is 36.1\% more accurate than the best of the previous techniques on average. By providing comprehensive architecture, SARIF can help users understand systems effectively and reduce the manual effort of obtaining ground-truth architectures.

\end{abstract}

\maketitle

\section{Introduction}\label{sec_intro}

Software systems change over time to include new features or for maintenance purposes. During this process, actual implementation of the system may become different from the initial architectural design, which is known as the architectural \textit{drift} and \textit{erosion}~\cite{medvidovic2010software}. Architectural \textit{drift} and \textit{erosion} can lead to wrong decisions on development activities that result in a waste of time and development resources.
In order to prevent such \textit{drift} or \textit{erosion}, architects need to understand the implemented architecture. However, such task is costly, which can take hundreds of hours of an expert for a system with hundred thousands of lines of code~\cite{garcia2013obtaining}.

To reduce the manual effort required to maintain the architecture, various techniques have been proposed to recover the system architecture from its code implementation. These techniques aim to associate software implementation-level entities (i.e., files, functions or classes) with high-level system components (i.e., clusters of entities) by clustering the entities according to their structural dependencies or textual information.
For example, ACDC~\cite{tzerpos2000accd}, Bunch~\cite{mancoridis1999bunch}, and FCA~\cite{teymourian_fast_2022} cluster the entities based on the structural dependencies of software systems. 
Architecture Recovery using Concerns (ARC)~\cite{garcia2011enhancing} mostly relies on the semantic, i.e., topic modeling results of the source codes, to cluster the entities.
SADE~\cite{papachristou_software_2019} clusters the software system by combining both the dependency-based and textual-based results.
However, according to  \cite{lutellier2015comparing}, 
the current results of architecture recovery techniques are not accurate enough for real-world applications. 
We summarize the two major limitations of the existing solutions as follows.


First, the existing research recovers the architecture based on only one or two types of information to group software entities, despite a rich set of information can be helpful. For example, dependency between entities is the most commonly used information. In addition to dependency, ARC~\cite{garcia2011enhancing} has proven that textual information (i.e. information from code text) is also useful for architecture recovery.
Moreover, package or folder structures of software systems also carry important structural information on the architecture~\cite{kobayashi2012feature}.
Techniques focusing on only one or two types of information rather than all these aspects may produce one-sided and inaccurate recovery results of architecture.

Second, previous research tends to use coarse-grained information, which lacks the details that are critical to architecture recovery.
For example, in dependency-based works \cite{tzerpos2000accd, mancoridis1999bunch, teymourian_fast_2022}, all the dependencies are treated to have equal contributions to the clustering. 
However, different types of dependencies have different influences on the architecture. According to~\cite{stavropoulou2017case}, \textit{Function Call} dependencies can reflect much more architectural information than \textit{Use Type} dependencies.
Moreover, dependencies introduced by different entities also have different impact on the architecture. For example, dependencies between public interfaces of modules are more important than that between internal implementations~\cite{bass2003software}.

\begin{figure}
    \centering
    \fbox{\includegraphics[width=0.3\textwidth]{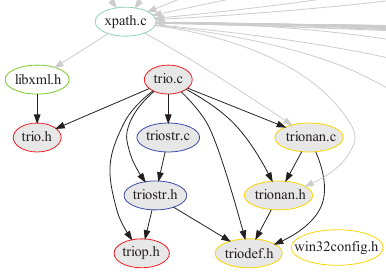}}
    \caption{\textit{trio} Module in Libxml2}
    \label{fig_libxml2}
\end{figure}

To overcome these limitations, we propose \ourtool, a fully automated architecture recovery technique that fuses dependency, code text and folder structure information to produce comprehensive recovery results.
\ourtool is a cross-platform architecture recovery tool that currently supports C, C++ and Java. 
To obtain fine-grained information, \ourtool 
1) refines the entity-level dependencies with entity importance and dependency type;
2) measures the strength of file-level dependencies based on the entity-level ones; 
3) uses TF-IDF and LDA to focus on topic information instead of raw code text;
4) selects folder structures based on their quality;
Last, \ourtool builds an adaptive model to fuse different types of information and clusters the software entities to recover the architecture.


We evaluate \tool with established  metrics MoJoFM~\cite{wen2004effectiveness}, a2a~\cite{le2015empirical}, \ccg~\cite{le2015empirical} as well as two additional metrics, Adjusted Rand Index (ARI)~\cite{hubert1985comparing} and \aaj, as we notice that the commonly used architecture similarity metrics may not be adequate to produce unbiased architecture comparison results. 
We compare \ourtool against 7 previous techniques on 9 software systems with human-labeled ground-truth architecture.
The experiments show that \ourtool outperforms previous architecture recovery methods by being \mj{36.1\%} more accurate than the best of the previous results on average. Moreover, we did an extensive experiment on 900 GitHub projects to further verify the generalizability of \ourtool.

In summary, our main contribution includes:
\begin{itemize}
    \item We propose \ourtool, a software architecture recovery algorithm that leverages the fine-grained program dependencies, code text and folder structure to produce comprehensive recovery results.
    \item We introduce 
    ARI to the scenario of comparing architectures, and design a new metric \aaj which solves limitations of $a2a$ to measure the architecture similarities.
    \item We evaluate \ourtool with 9 existing techniques on 9 software systems with ground truth architectures. The results show that \ourtool outperforms other techniques by being \mj{36.1\%} more accurate on average.
\end{itemize}


\section{Motivating Example}\label{sec_background}



In this section, we illustrate our motivation with some examples of architecture recovery results of existing techniques. We collaborated with a big software vendor company (name anonymized) to manually label the architecture of \textit{Libxml2}~\cite{repo-libxml2} (v2.4.22), an XML toolkit in C. Then, we compared the architectures recovered by state-of-the-art architecture recovery tools with our manual labeling, and found that the recovery results were unsatisfactory.

For example, we manually checked their result on the \textit{trio} module of
\textit{Libxml2}. The \textit{trio} module is a portable implementation of \textit{printf} function family. It contains 8 files which are marked with gray background in Fig~\ref{fig_libxml2}. The \textit{trio} module is one of the most distinctive module in \textit{Libxml2} from both structural and textual perspectives. Structurally, 14 out of 17 file dependencies are internal dependencies. Textually, 1811 out of 1857 \textit{trio} words in \textit{Libxml2} are from this module. However, most of the state-of-the-art architecture recovery tools still failed to recognize this module by dividing it into multiple  modules or combining them with other modules. We manually analyze the results of two tool, ARC and Bunch,  and summarize the reasons for their failures as follows.


ARC is a text-based architecture recovery tool. Its clustering result is shown in Fig~\ref{fig_libxml2} by the color-coding of nodes' frames. 
ARC divides the \textit{trio} module into three clusters. Two of them (red and blue) only comprise files from the \textit{trio} module. 
However, the third cluster (yellow)  includes 3 files from \textit{trio} and \textit{win32config.h}, which is a configuration file and not really related to \textit{trio} module. This is probably because \textit{win32config.h} contains the most \textit{trio} words outside the \textit{trio} module (11 out of 46), even though it has completely different functionalities from the \textit{trio} module.
This demonstrates that solely use textual information may be inadequate since the code text does not necessarily reflect the code's functionality.

{Bunch recovers the architecture based only on dependency information. It fails to identify the \textit{trio} module and clusters}
these 8 files together with another 31 files. 
For example, \textit{xpath.c} is the gateway of its \textit{xpath} module, and many other modules communicate with this module via \textit{xpath.c}. Especially, the \textit{trio} module has the closest dependencies related to \textit{xpath.c}, which misleads Bunch to cluster \textit{xpath.c} into the \textit{trio} module. However, according to the dependency analysis result of the \textit{Depends} tool~\cite{multilang-depends}, only less than 1\% (13/1359) of entity-level dependencies related to \textit{xpath.c} are related to the \textit{trio} module. This indicates that although dependency information is proven to be useful in architecture recovery, inadequate usage
could also bring biases to the architecture recovery. 

\ourtool fuses various aspects of information from the source codes, including both dependency and textual information, while using them in a fine-grained manner to cluster the software components more precisely. According to our experiment, the \textit{trio} module can be perfectly identified by \ourtool, and the recovered result is much more similar to the ground truth compared to Bunch and ARC.

\section{Methodology}\label{sec_method}


\begin{figure}[tb]
    \centering
    \includegraphics[width=0.47\textwidth]{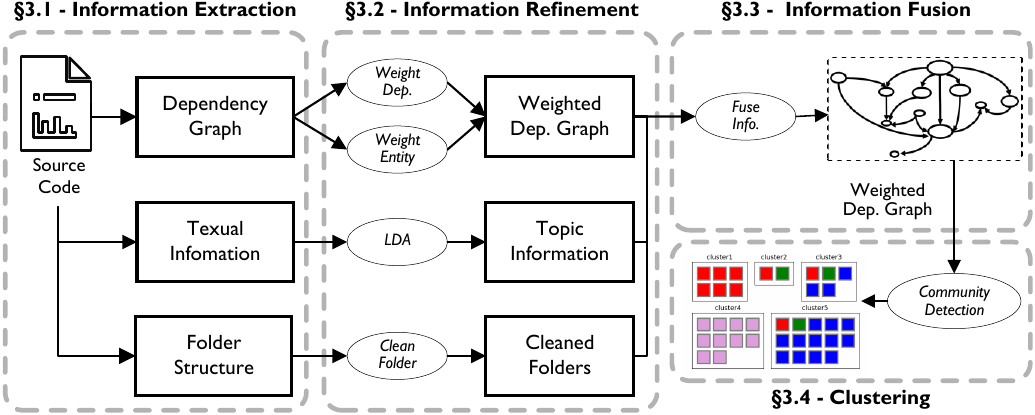}
    \caption{Overview of Our Approach}
    \label{fig_method_overview}
\end{figure}

The overview of \ourtool is presented in Fig~\ref{fig_method_overview}. 
\ourtool first extracts three types of information from the given software. Then, it refines the information to obtain fine-grained information with the help of program analysis and machine learning techniques. Next, it fuses the information into one weighted graph to represent relations of the entire project.
Last, based on the graph, \ourtool clusters the system into different groups to recover the architecture.

\subsection{Information Extraction}
In this section, we will introduce the details of how to extract the three aspects of information, i.e., dependencies, textual information and folder structure. 

\subsubsection{Dependencies}\label{subsubsec_dep_extr}
The dependency is the most important information for recovering the architecture of software systems, as it contains all the structural interactions in the software system.
We use a dependency graph to represent the dependencies between entities in software systems. The nodes of the graph represent software entities (e.g. file, class, method/function, variable, etc.), and the edges represent the relation between two entities (e.g. function call, file import, etc.).
\ourtool uses \textit{Depends}~\cite{multilang-depends} to generate dependencies for target software systems. \textit{Depends} generates dependency information by observing the syntactic structures among software elements in C/C++/Java projects. 
It can detect entities include functions/methods, variables, type definitions, etc, along with 13 types of relations between these entities.
In this sense, the dependency model of \textit{Depends} can tell exactly the categories of both the dependency relation and the nodes. This detailed information allows us to generate more accurate dependency graphs of software systems, while previous studies 
suggest that more accurate dependencies generally result in better architecture recovery results~\cite{lutellier2015comparing, lutellier2017measuring}.


\subsubsection{Text from Source Code}
Textual information can help the recovery since the developers embed their domain knowledge into code texts by means of comments, names of methods, classes, etc. For example, as discussed in Sec~\ref{sec_background}, \textit{trio} module contains 1811 word \textit{trio} out of a total of 1857 in the entire system, which is a strong indication of the boudary of the \textit{trio} module.

The textual information of a software system  mainly comes from two sources: the source code and the documentation. \ourtool mainly focuses on the texts in the source code since \ourtool is a fully automated tool and it is hard to automatically match the documentation to their corresponding codes.

\ourtool uses Ctags~\cite{github-ctags} and comment parser~\cite{pypi-comment-parser} to extract words from different fields of functions and comments.
To reduce the noise introduced by unimportant words, only the following kinds of words will be extracted: 1) filename, 2) definitions of classes, functions/methods, and global variables, 3) comments. 

\subsubsection{Folder Structure}\label{subsubsec_ext_fold}
The folder structure shows how the developers organize the files, which gives hints to recover its architecture. 
Therefore, it is included as one of our information sources. However, not all of the folder structures can provide positive suggestions for the architecture recovery~\cite{garcia2013obtaining}.
For example, many C/C++ projects organize their header files into a separate folder which provides no insight into the architecture. Moreover, some folders like the 'mapred/lib' folder of \textit{Hadoop} (v0.2) contains multiple libraries for various purpose. However, these libraries actually belong to multiple independent submodules. As a result, grouping all the files within this folder together is not reasonable.
We will address how to eliminate these  improper folder structures in the next step. 


\subsection{Information Refinement}

\subsubsection{Dependency Graph}\label{subsubsec_process_dep}
In this step, we start with the unweighted dependency graph extracted in Sec~\ref{subsubsec_dep_extr}. 
To make it reflect the architecture more accurately, we apply two types of weights to the graph. 
The first weighting is for entities, i.e., the nodes in the dependency graph, since different entities have different degrees of impact on architecture. For example, interfaces of modules have more influence on the architecture than the modules' internal implementations~\cite{bass2003software}, thus they should have higher weights.
The second weighting is for dependencies (i.e., the edges in the dependency graph), since different types of dependencies carry different amounts of architectural information~\cite{stavropoulou2017case}.


\noindent\textbf{Weigh Entities by Importance.} The dependency graph comprises different types of entities, e.g. files, functions, variables, etc. 
It is unfair to evaluate the importance of all entities directly since their granularity is not comparable. Therefore, we start with evaluating the function-level entities, then derive the weight or entities at other granularities based on it. The reason for prioritizing function-level entities is that entities at the variable level have limited information, and entities at the class/file level are too coarse-grained.

\mj{To measure node importance in a graph, a widely used algorithm is PageRank~\cite{brin1998anatomy}.
PageRank evaluates a node's importance by its popularity, i.e. a node is important if it can be easily accessed by many other nodes. However, in the dependency graph, PageRank may overvalue utility functions like parsers.
For example, in \textit{Libxml2}, the most important function identified by PageRank is \textit{xmlStrEqual}, which is a utility function that compares the strings in XML. However, this function has nearly no indication on the architecture since it is accessed by almost every module in \textit{Libxml2}. Conversely, interfaces, which are impactful on the architecture~\cite{bass2003software}, tend to be less accessible thus undervalued. This scenario contradicts our starting point that interfaces are more influential on architecture than internal implementations like utilities. To address this, we use Inversed PageRank (IPR)~\cite{gyongyi2004combating}, which is a variant of PageRank, to determine the importance of each function. Different from PageRank, IPR is designed to find the influential nodes, i.e., the nodes that can access as many as other nodes. Intuitively, module interfaces have access to the module's internal implementations, thus will be assigned with higher importance. 
For example, in \textit{Libxml2}, the functions with top 2 importance are: 1) \textit{main} function of \textit{xmllint.c}, which is the top interface of the \textit{xmllint} module. 2) \textit{xmlXPathNewContext}, which is one of the most commonly-used API of the \textit{xpath} module.}

\mj{To be more detailed, IPR assigns importance to the functions in two ways. The first portion of importance is assigned based on the count and quality of a function's callees. This method iteratively computes the IPR values by propagating importance from callees to their callers until the results converge.
The second portion of importance is equally spread across all nodes. A specific parameter, known as the damping factor $d$, decides the distribution of overall importance between these two sections. Mathematically, IPR is defined by the following equation:}
\begin{equation}\label{eq_ipr}
    IPR(n_i) = d\sum_{n_i \in P(n_i)}\frac{IPR(n_j)}{in\_degree(n_j)} + \frac{1-d}{N}
\end{equation}
where $n_1\sim n_N$ are nodes in graph, $N$ is the total number of nodes, $P(n_i)$ are the direct successors of $n_i$, $d$ is the damping factor. \mj{In \ourtool, $d$ is set to 0.85, which follows the recommendation of the algorithm's authors~\cite{brin1998anatomy}. The result of IPR ranges from 0 to 1.}

To assess the importance of the functions, a function dependency graph is extracted from the dependency graph by keeping only the function-level entities and dependencies. The importance of a function is then defined as its IPR result.
Based on the importance of functions, the importance of higher-level entities (e.g. file, class) is defined as the sum of the functions they contained. \mj{The reasoning behind using summation is that the dependencies associated with these higher-level entities are considerably fewer than those related to functions. Utilizing the sum is to guarantee that these higher-level entities, which are also crucial for the architecture, will not be overlooked due to their lesser dependency count.}
\mj{On the other hand, the importance of lower-level entities is determined by evenly dividing the significance of the parent entity amongst them to prevent them from overshadowing the other entities.}




\noindent\textbf{Weigh Dependencies by Type.} 
Entity-level dependency has many different types, 
and different types of dependencies have different level of indication on the architecture. For example, according to the previous study~\cite{stavropoulou2017case}, \textit{Function Call} dependencies can reflect much more architectural information than \textit{Uses Type} (i.e., a function accesses a specific type) dependencies. 
However, this limited understanding of their relative usefulness in architecture is not enough for us to manually assign appropriate weights to all 13 types of dependencies extracted by \textit{Depends}.


Therefore, we designed an algorithm to automatically assign reasonable weights to each type of dependency. 
Since the target of this weighting is to make the dependency graph reflect more architectural information, our algorithm optimizes the weights to make the weighted dependency graph can be clustered into a better architecture.
Specifically, our algorithm consists of 5 steps:
1) Randomly set the weights of different types of dependencies;
2) Extract the unweighted dependency graph, then weigh the edges based on step 1;
3) Cluster the weighted graph using community detection algorithm~\cite{clauset2004finding};
4) Calculate the Girvan-Newman modularity quality (MQ)~\cite{reichardt2006statistical} of the clustering result, where MQ is a commonly used metric for measuring the quality of a clustering result;
5) Iterate step 1 to 4 to find the weight that can maximize the modularity.


To get a reasonable result from our algorithms, we collected 100 project for each of the languages supported by \ourtool (i.e., C, C++, Java). The projects are collected by searching for popular projects on GitHub to ensure their quality, and the projects that will be used in the following evaluation sections have been excluded to prevent presenting overfitting results.
\mj{The algorithm is implemented with \textit{hyperopt}~\cite{hyperopt-repo} optimizer.
The weight of each type is limited between 0.1 and 10, \mj{the converge condition is set to when the loss is less than 1e-5,} and the optimization results are shown in Table~\ref{tab_dep_weights}. The resulting weights basically align with our intuitive expectation and the previous~\cite{stavropoulou2017case}. For example, \textit{ImplLink} (similar to \textit{Function Call}) is assigned a much higher weight than \textit{Use} (similar to \textit{Uses Type}). 
This result will be combined with the entity importance to form a weighted dependency graph.}

\begin{table}[tb]
    \caption{The Optimized Weights of Different Types of Dependencies. "MixIn" Type is not Applicable to C/C++/Java.}\footnotesize
    \begin{tabular}{rrrrrrr}
        \hline
        Implement & Throw & Call & Create & ImplLink & Extend & Use \\ 
        7.575 & 9.053 & 0.177 & 1.665 & 9.33 & 2.159 & 0.509 \\ \hline
        Parameter & Import & Cast & Return & Contain & MixIn &  \\ 
        5.146 & 8.300 & 0.701 & 5.702 & 4.478 & N.A. &  \\ \hline 
    \end{tabular}
    \label{tab_dep_weights}
\end{table}

\noindent\textbf{Merge Weight Results.}
Based on both the weight of dependency type and related entities, the final dependency weight is defined as:
\begin{equation}\label{eq_dep}
    Weight(e) = type\_weight(e)\cdot \frac{impt(src)+impt(dst)}{2}
\end{equation}
\mj{where $type\_weight(e)$ is defined in Table~\ref{tab_dep_weights}, $impt(src)$ and $impt(dst)$ are the importance of the dependency's source and target entity in Eq~\ref{eq_ipr}, respectively.} A file-level dependency graph is then formed by grouping all entities into it
s corresponding files. The weight of dependency between two files is defined as the sum of its corresponding entity-level dependencies.

\subsubsection{Textual Information}\label{subsubsec_process_text}
The refinement of textual information comprises three steps: preprocessing, weighing and topic modeling.

\noindent\textbf{Preprocessing.} \ourtool preprocesses the extracted raw text by splitting the terms according to common naming conventions of programs, such as snake-case or camel-case. For example, the function name "plotFigure" will be split into "plot" and "figure". Then, \ourtool applies commonly-used preprocessing techniques including lemmatization and stop word removal to refine the words.


\noindent\textbf{Weighing.} \ourtool assigns different weights to the words based on three criteria: 1) 
The importance of texts extracted from different parts of the source code is not equivalent. 
For example, a word in comments is likely to be less important than that in function declarations. Thus, the texts will be weighted according to where they come from;
2) A word from a less important entity is less likely to be useful. Therefore, \ourtool modifies the weights of words according to its source entity as defined in Sec~\ref{subsubsec_process_dep};
3) \ourtool uses TF-IDF~\cite{salton1988term} to decrease the weight of common words.
The final weight of each word will be the product of the three weights:
\begin{equation}
    Weight(w) = Weight_{src}(w)\cdot Weight_{entity}(w)\cdot TF-IDF(w)
\end{equation}

\noindent\textbf{Topic Modeling.} \ourtool summarizes the words from each file into a topic via Latent Dirichlet Allocation (LDA)~\cite{blei2003latent}. Specifically, it trains a LDA model with the weighted words. Each source file is assigned with a topic embedding using the trained model. Last, \ourtool calculates the correlation between files on the topic embeddings. 

\subsubsection{Folder Structure}\label{subsubsec_process_pack}
In this step, we aim to filter out folders that most likely to be useless. 
As discussed in Sec~\ref{subsubsec_ext_fold}, most of the folders that have no indication on the architecture contain files from multiple modules. These files have little dependencies with each other but are strongly related to files outside the folder. \mj{Based on this characteristic, our filtering algorithm is designed to remove folders that have high inter-dependencies and low inner-dependencies. Specifically, if the folder has more inter-dependencies than the inner-dependencies, it will be eliminated by being merged into its parent folder.}

\subsection{Information Fusion}\label{subsec_info_fuse}
Based on the processed information, \ourtool fuses them to have a more comprehensive understanding of the system architecture. To this end, \ourtool first evaluates the quality of each type of information, and assigns weight accordingly. Then, \ourtool fuses the information into a unified graph.


\subsubsection{Weight Assignment}\label{subsubsec_weight}
The three types of information can reveal different degrees of ground truth architecture from different perspectives. Intuitively, the information with higher quality (i.e., can produce results closer to the ground truth architecture) should be assigned a higher weight. Thus, we need to estimate the quality of the three types of information, then assign weights accordingly.


Among the three information, the dependency information represents the relations between software entities, which objectively reveal the interactions among the modules. 
Such characteristics ensures that the dependencies are highly likely to have the capacity to reveal an adequate portion of the architecture. However, the quality of textual information and folder structure is not as objective as the dependencies as they are produced by the developers and inherently subjuctive.
Therefore, their quality could vary greatly in different software projects. 
For instance, recall the motivating example in Sec~\ref{sec_background}, the \textit{trio} module in \textit{Libxml2} has strong textual characteristics so that the textual information can be a good indicator to distinguish modules. However, all the source files of \textit{Libxml2} are located in its root folder, which means that the folder structure is useless in the architecture recovery. 
Moreover, in other software projects, the situation could be completely different. The \textit{builtin} module of \textit{Bash-4.2}, for example, contains implementations of built-in commands like \textit{cd}, \textit{echo}, etc. Due to the wide variety of commands within the module, its textual information is overly general, which reduces its usefulness in recovering the architecture. Consequently,
text-based techniques such as ARC struggle in identifying the \textit{builtin} module by grouping its files with those having similar text feature from other modules. 
Despite the \textit{builtin} module lacks distinctive textual features, its content is identical to that in the \textit{builtin} folder, which means the module can be easily identified by the folder structure.

To handle such extreme variance in the quality of textual information and folder structure, a dynamic weight is assigned to them by comparing the similarity of the architecture recovered through these information with the architecture recovered through dependencies. 
The underlying reason for such an assignment is that the architecture recovered through dependencies is expected to have some similarity with the ground truth due to the inherent relations between dependencies and architecture.
\zz{For textual information or folder structure, if its quality is high, the recovered architecture based on it should have a higher similarity with the ground truth, and therefore more likely to have higher similarity with the architecture recovered through dependencies.}


\begin{table}[tb]
    \caption{The weight of information of \textit{Libxml2} and \textit{Bash}.}\small
    \begin{tabular}{c|cc}
        \hline
        Project & textual information & folder structure\\
        \hline
        \textit{Libxml2} & 0.639 & 0.068\\
        \textit{Bash} & 0.141 & 0.703\\
        \hline

    \end{tabular}
    \label{tab_sample_weights}
\end{table}

To implement this design, \tool first recover the architecture based on only one type of information.
The ways \ourtool recover architecture with each type of information solely are as follows:

\begin{itemize}[leftmargin=*]
    \item Dependency: The weighted dependency graph in Sec~\ref{subsubsec_process_dep} is utilized for clustering, in which \tool groups the files through the community detection algorithm~\cite{clauset2004finding}.
    \item Textual: An undirected weighted graph is generated based on the topic correlations between files from Sec~\ref{subsubsec_process_text}. The nodes in the graph represent files, and the edges are weighted according to the topic correlation between the two files. The graph is then clustered by the hierarchical clustering algorithm with the complete linkage metric~\cite{naseem2013cooperative}.
    \item Folder: The clustering result is formed by the filtered folder structure from Sec~\ref{subsubsec_process_pack}.
\end{itemize}

\mj{Next, the similarities between different types of information are measured based on the \aaj similarity of the recovered architectures, where \aaj will be detailed  in Sec~\ref{subsec_new_metric}. The weight of each supporting information is defined as}:
\mj{
\begin{equation}\label{eq_weight_fusion}
    Weight_s = a2a_{adj}(A_{dep}, A_{s})
\end{equation}
}
\mj{where $s$ refers to textual information of folder information, $A_s$ is the recovered architecture of the corresponding information.}
Our method is tested on \textit{Libxml2} and \textit{Bash} to confirm its ability to detect differences in the quality of textual information and folder structure. The results of the weight assignment for textual information and folder structure are presented in Table~\ref{tab_sample_weights}, which match well with our manual inspection of the two projects.



\subsubsection{Textual Information Integration}
Based on the weight of textual information, the weight of edges is adjusted according to the textual correlations between the corresponding files. 
All existing edges are modified with a coefficient $coef_t$:
\mj{
\begin{equation}
    coef_t = 1 + corr*Weight_{text}
\end{equation}
}
where $corr$ is the correlation of topic embedding between the two files on the edge, and $Weight_{text}$ is the weight of textual information we assigned in Eq~\ref{eq_weight_fusion}. 
Moreover, the purpose of applying $coef_t$ is to increase the weight of dependency between two files if their textual features are similar, and vice versa. 
To this end, if $coef_t$ between two files are greater than 0.8, which means their textual features are highly similar, a bi-directional edge will be added between the files even if they do not have structural dependencies.


\subsubsection{Folder Structure Integration}
As the files in the filtered folders are more likely to belong to the same modules, the weight of edges between files that are located in the same folder are increased by:
\mj{
\begin{equation}
    coef_f = \frac{1}{1 - Weight_{folder}}
\end{equation}
}
where $Weight_{folder}$ is the folder weight defined in Eq~\ref{eq_weight_fusion}. \mj{$Coef_f$ ranges from 1 to +$\infty$. 
This coefficient is designed to enhance relations between files in the same high-quality folder since a well-structured folder is a strong indicator of a module.
}



\subsubsection{Generate Final Graph} 
We generate the final graph by updating the edge weights of the input dependency graph from Sec~\ref{subsubsec_process_dep}:
\begin{equation}
    Weight_{final} = Weight_{ori} \cdot coef_t \cdot coef_f
\end{equation}
where $Weight_{ori}$ is the weight defined in Eq~\ref{eq_dep}. Thus far, the graph carrying all types of information is generated by updating the weights and is ready for clustering.

\subsection{Clustering the Final Graph}\label{subsec_final_cluster}
\begin{figure}
    \centering
    \includegraphics[width=0.33\textwidth]{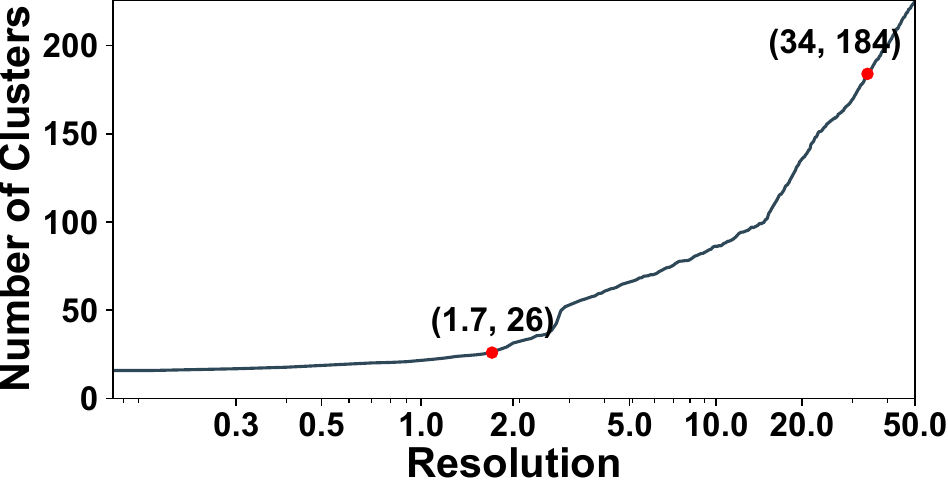}
    \caption{\mj{Number of Cluster v.s. Resolution}}
    \label{fig_resolution}
\end{figure}

In this final step, we cluster the graph that carries the fused information into modules. 
We use the community detection algorithm proposed by Clauset et al.~\cite{clauset2004finding} for clustering. The algorithm cluster a graph by maximizing its modularity which is defined as:

\begin{equation}\label{eq_gn_mq}
    Q_{\gamma}=\frac{1}{2m} \sum_{i, j}\left[A_{i j}-\gamma\frac{k_{i} k_{j}}{2 m}\right] \delta\left(C_{i}, C_{j}\right)
\end{equation}
where $A_{ij}$ is the weight of edge from node $i$ to $j$; $k_i$ denote the sum of weights of edges incident on $i$; m is the total weight of edges; $\gamma$ is a parameter also known as \textit{resolution}, which affect $Q_{\gamma}$'s preference on cluster size.
The reason we choose this community detection algorithm is that it runs much faster than most of the existing clustering algorithms like the Bunch search algorithm, especially when the software system is large.
\mj{The granularity, i.e., the number of clusters, of the final architecture can be controlled by adjusting the \textit{resolution} parameter. 
Since different granularity of architecture might be desired for different use cases, no universal standard can be established for the granularity of architecture. Therefore, the \textit{resolution} of \ourtool is user-controllable, and general correlation between the number of clusters and the \textit{resolution} is shown in Fig~\ref{fig_resolution} for user reference. 
For users unsure about what level of granularity to choose, we suggest 20 to 30 clusters as a suitable setup for quickly understanding an architecture. Accordingly, the default value of \textit{resolution} is set to 1.7.}

\section{Similarity Metrics}\label{sec_metric}


To evaluate architecture recovery results, a common method is to compare the result with human-labeled ground truth architectures. In previous studies, the three most commonly used metrics are MoJoFM, a2a, and \ccg. However, each of these metrics has its own limitations. In this section, we will discuss these limitations, and introduce two new metrics to address them.

\subsection{Previous Metrics and their Limitations}\label{subsec_pre_metrics}

\noindent \textbf{MoJoFM}~\cite{wen2004effectiveness} is defined by the following equation:
\begin{equation}
    MoJoFM(M)=\left(1-\frac{mno(A, B)}{\max (mno(\forall A, B))}\right) \times 100 \%
\end{equation}
where $mno(A, B)$ is the minimum number of move or join operations needed to transform an architecture $A$ into $B$. 
MoJoFM is the most commonly used metric for comparing architectures. However, MoJoFM prefers architectures with many small clusters~\cite{lutellier2017measuring} due to its low cost for merging two clusters.


\noindent \textbf{Architecture-to-architecture} ($a2a$)~\cite{le2015empirical} 
is a distance-based measurement, which is defined as:

\begin{equation}
    a2a(A, B)=\left(1-\frac{mto(A, B)}{aco(A)+aco(B)}\right) \times 100 \%
\end{equation}
where $mto(A, B)$ is the minimum number of operations needed to transform architecture $A$ into $B$, $aco(A)$ is the number of operations needed to construct architecture $A$ from a null architecture. $a2a$ has nearly no preference over the number of clusters.
However, studies show that it has a limited variations~\cite{lutellier2015comparing}, which means an architecture that greatly differs from the ground truth may not receive a low score as it deserves.

\noindent \textbf{\boldsymbol{$c2c_{cvg}$}}~\cite{le2015empirical} is a metric defined by the number of similar clusters between two architectures:
\begin{equation}
    c2c_{cvg}(A, B)=\frac{\left|simC(A, B)\right|}{\left|A \right|} \times 100 \%
\end{equation}
where $\left|A \right|$ is the number of clusters in $A$; $simC(A, B)$ is the number of similar clusters between $A$ and $B$, where two clusters are defined to be similar if their overlapping percentage passed the user-defined threshold $th_{cvg}$.  
According to our experiments which will be detailed in the following section, \ccg's performance could be unstable in many scenarios.




\subsection{New Metrics}\label{subsec_new_metric}
To address the limitations of previous metrics, two additional metrics are introduced in this paper.


\noindent \textbf{Adjusted Rand Index (ARI)}~\cite{hubert1985comparing} is a well-known metric used to quantify the degree of similarity between two distinct partitionings of a single entity set. The versatility of ARI allows it to be applied in comparing two architectures of the same system. In general, the ARI computes the proportion of entity pairs that are consistently clustered in both architectures. The higher proportion of consistent pairs, the more similar the two partitionings are.
Mathematically, ARI is calculated as follows:

\begin{equation}
ARI=\frac
{\sum_{ij}\binom{n_{ij}}{2}-\left[\sum_{i}\binom{a_{i}}{2} \sum_{j}\binom{b_{j}}{2}\right] /\binom{n}{2}}
{\frac{1}{2}\left[\sum_{i}\binom{a_{i}}{2}+\sum_{j}\binom{b_{j}}{2}\right]-\left[\sum_{i}\binom{a_{i}}{2} \sum_{j}\binom{b_{j}}{2}\right] /\binom{n}{2}}
\end{equation}

In this formula, $a_i$ and $b_j$ correspond to the total count of entities present in the $i$-th cluster of architecture $A$ and the $j$-th cluster of architecture $B$, respectively. $n_{ij}$ is the count of entities that are both in the $i$-th cluster of architecture $A$ and the $j$-th cluster of architecture $B$. $ARI$ is widely recognized and extensively tested, exhibiting no evident drawbacks such as a low dynamic range or a bias towards cluster size. Our evaluation which will be detailed in the following section further proved its effectiveness.

\noindent \boldsymbol{$a2a_{adj}$} is designed by us to address the issue of limited variation observed in the $a2a$ measure. The limited variation in $a2a$ originates from the excessively large denominator, which is the sum of the architecture costs $aco(A)+aco(B)$, used in the calculation of $mto$.
To alleviate this issue, we decompose the distance (i.e., $mto$) of $a2a$ into two distinct components: 1) $mto_m$, which corresponds to the cost of reassigning entities to a different cluster; and 2) $mto_{ar}$, reflecting the cost of adding or removing entities/clusters. Our adjusted measure \aaj is then defined as follows:


{
\setlength{\tabcolsep}{1.85pt}
\begin{table}[tb]
    \caption{Information of Projects in Evaluation.}
    \label{tab_projects}\footnotesize
    \centering
    \begin{tabular}{lllllll}
        \hline 
        Name & Version & \mj{Language} & NLOC & File & Cluster & \mj{Domain}\\
        \hline \hline
        ArchStudio & 4  & \mj{Java} & 238K & 2305 & 57 & \mj{IDE}\\
        Bash & 4.2  & \mj{C} & 115K & 405 & 14 & \mj{Shell}\\
        Chromium & svn-171054 & \mj{C++} & 11.7M & 46,498 & 67 & \mj{Browser} \\
        Hadoop & 0.19.0  & \mj{Java} & 225K & 1703 & 67 & \mj{Distributed Framework} \\
        ITK & 4.5.2 & \mj{C++} & 1.23M & 8,504 & 11 & \mj{Image Processing}\\
        OODT & 0.2  & \mj{Java} & 81.4K & 892 & 216 & \mj{Data Management}\\
        \hline 
        Libxml2  & 2.4.22 & \mj{C} & 81.1K & 82 & 17 & \mj{XML Parser} \\
        HDC & 46ff87  & \mj{C++} & 25.7K & 207 & 11 & \mj{Camera Interface} \\
        HDF & 0e196f  & \mj{C} & 153K & 1,051 & 15 & \mj{Driver Subsystem}\\
        \hline
    \end{tabular}
\end{table}
}

\begin{equation} \label{eq_a2a_adj}
    \begin{split}
        a2a_{adj} &= 1 -
                \alpha\cdot\frac{mto_m(A,B)}{mto_{m}^{max}(A,B)} - 
                \beta\cdot\frac{mto_{ar}(A,B)}{aco(A)+aco(B)}, \\
        where \text{ } \alpha &= \frac{n_{shared}+min(nc_A, nc_B)}{n_{shared}+n_{diff}+max(nc_A, nc_B)}, \\
        \beta &= \frac{n_{diff}+abs(nc_A-nc_B)}{n_{shared}+n_{diff}+max(nc_A, nc_B)}
    \end{split}
\end{equation}

$n_{shared}$ and $n_{diff}$ denote the number of shared and unshared elements, respectively. $nc_A$ and $nc_B$ are the counts of clusters in architecture $A$ and $B$.
The measure $a2a_{adj}$ quantifies the dissimilarity between two architectures with two costs: 1) the cost of reassignment of shared entities to different clusters; 2) the cost of an addition or removal of entities or clusters. Each of the two costs is individually normalized by their appropriate denominators. The reassignment cost $mto_m$ is normalized by the maximum possible reassignments $mto_{m}^{max}$ given the number of entities and clusters, as solved in~\cite{charon2006maximum}. The addition/removal cost $mto_{ar}$ is normalized by the cost of building both architectures from scratch.

In the final step, the costs associated with these two aspects are weighted according to the number of shared and unshared components. By normalizing the two costs using a more appropriate denominator, the resulting dynamic range of \aaj is significantly larger than that of the original $a2a$ metric. This expanded dynamic range will be further proved in the evaluation section.

\section{Evaluation}\label{sec_eval}
To evaluate \ourtool, we aim to answer the following RQs:

\noindent
\textbf{RQ1}: What is the performance of the proposed similarity metrics in evaluating architecture recovery accuracy?

\noindent
\textbf{RQ2}: What is the accuracy of \ourtool in architecture recovery compared to related works?

\noindent
\textbf{RQ3}: What is the contribution to the architecture recovery for each type of information and weighting?








\subsection{Experimental Setup}\label{sec:expset}


\subsubsection{Baseline Selection}
Our selection of baseline techniques consists of two parts. The first part comprises 5 tools who show the most promising results in the previous empirical studies~\cite{garcia2013comparative, lutellier2015comparing, lutellier2017measuring}, including  ACDC, Bunch, ARC, WCA and LIMBO. 
\textbf{ACDC}~\cite{tzerpos2000accd} is a pattern-based clustering algorithm. 
\textbf{Bunch}~\cite{mancoridis1999bunch} is a search-based technique that recovers architecture by optimizing TurboMQ. 
\textbf{ARC}~\cite{garcia2011enhancing} recovers architecture with topic information rather than dependency. 
\textbf{WCA}~\cite{maqbool2004weighted} and \textbf{LIMBO}~\cite{andritsos2004limbo} are two hierarchical clustering algorithms that can be applied to architecture recovery. 


The second part contains the latest research on architecture recovery. We reviewed a total of 39 papers related to architecture recovery technique in the recent five years~\cite{cho_software_2019, huang_multi-agent_2017,link_value_2019, sun_pso_2018, olsson_incremental_2022, izadkhah_information_2019, prajapati_software_2021, hwa_search-based_2017, tarchetti_dct_2020, tan_e-sc4r_2021, elyasi_hygar_2022, papachristou_software_2019, lutellier_measuring_2018, shatnawi_recovering_2017, zahid_evolution_2017, bi_systematic_2018, hoff_towards_2021, teymourian_fast_2022, garcia_constructing_2021, link_recover_2019, link_study_2021, schmitt_laser_arcade_2020, lee_identifying_2020, yang_enhancing_2022, boerstra_stronger_2022, kargar_multi-programming_2019, mohammadi_new_2019, psarras_mechanism_2019, amarjeet_many-objective_2018, rathee_improving_2018, ieva_discovering_2018, benkoczi_design_2018, singh_software_2017, lee_class_2017, li_framework_2017, kargar_semantic-based_2017, marian_hierarchical_2017, sadat2019multi, yano2020moderate}. 5 of 39 papers~\cite{teymourian_fast_2022, yang_enhancing_2022, boerstra_stronger_2022, psarras_mechanism_2019, papachristou_software_2019} contain link to their supplementary materials, and 4 of them~\cite{teymourian_fast_2022, yang_enhancing_2022, psarras_mechanism_2019, papachristou_software_2019} provide their artifact for reproduction. Among the 4 techniques, EVOL~\cite{yang_enhancing_2022} and CodeSum~\cite{psarras_mechanism_2019} take some intermediate data as input and we failed to generate such data for them since the data formats are not documented. Thus, the second part of baseline techniques comprises the remaining two techniques: FCA~\cite{teymourian_fast_2022} and SADE~\cite{papachristou_software_2019}.


\textbf{FCA}~\cite{teymourian_fast_2022} clusters software systems by performing operations on the dependency matrix. It has good scalability and can cluster very large software systems within a reasonable amount of time.
\textbf{SADE}~\cite{papachristou_software_2019} clusters software systems by combining the call graph with the semantic similarity between the user-defined modules.

\subsubsection{Baseline Implementation and Parameters}
We obtained the executable of ACDC, Bunch, FCA and SADE from the author's websites. For WCA, LIMBO and ARC, we adopted the implementations from ARCADE~\cite{schmitt2020arcade}. 
Since WCA, LIMBO and ARC allows user to select a preset number of clusters, we ran these tools with 10 to 100 clusters with an incremental step of 5. In addition, the ARC tool takes the number of concerns as input as well. We experimented ARC with 50 to 150 topics with a step of 10.

\subsubsection{Data Collection}\label{subsubsec_data_collection}
Our evaluation utilizes a two-part dataset. The first part includes projects with well-established architectures, which are used as a benchmark for assessing the accuracy of \ourtool and comparing \ourtool to the baselines. 6 open-source projects with ground-truth labeled in previous studies are included in this part: ArchStudio4, Bash-4.2, Chromium, Hadoop, ITK and OODT~\cite{garcia2013obtaining, lutellier2015comparing}. Moreover, 3 projects, Libxml2~\cite{repo-libxml2}, Distributed Camera~\cite{repo-dist-cam}, and Drivers Framework~\cite{repo-drv-frame}, have been labeled with our industry collaborators. Details of these projects are listed in Table~\ref{tab_projects}. The size of these projects, as measured by lines of code, ranges from about 20K to 11M, which is either equivalent to or exceeds the scale of projects used in previous studies~\cite{lutellier2017measuring, teymourian2020fast, elyasi_hygar_2022}. 
The second part comprises 900 popular projects from Github. The purpose of including this extensive dataset is to demonstrate the generalizability of \ourtool. To this end, we collected projects for each of the languages that supported \ourtool, i.e., C, C++ and Java. To avoid biased distribution of project size, we gathered projects that fell within three different size ranges: 0-10 MB, 10 MB-100 MB, and $\geq$100 MB. We used the GitHub API to search for repositories with the highest number of stars 
, and excluded the projects used in the parameter optimization detailed in Sec~\ref{subsubsec_process_dep}. For each size range and language combination, 100 projects are collected, resulting in a total of 900 projects.

\subsection{Metric Performance Evaluation (RQ1)}
In RQ1, we aim to evaluate the metrics on their ability to measure the similarity between architectures with two experiments.
All five metrics discussed in Sec~\ref{sec_metric} will be evaluated in this RQ: MoJoFM, a2a, $c2c_{cvg}$, ARI, and $a2a_{adj}$. Since $c2c_{cvg}$ has a threshold parameter, four different thresholds are tested following the settings of previous studies~\cite{lutellier2015comparing, le2015empirical}: 66\%, 50\%, 33\% and 10\%.




\begin{figure}[t]
    \centering    
    \subfigure[MoJoFM, a2a, ARI and $a2a_{adj}$]{
        \includegraphics[width=0.21\textwidth]{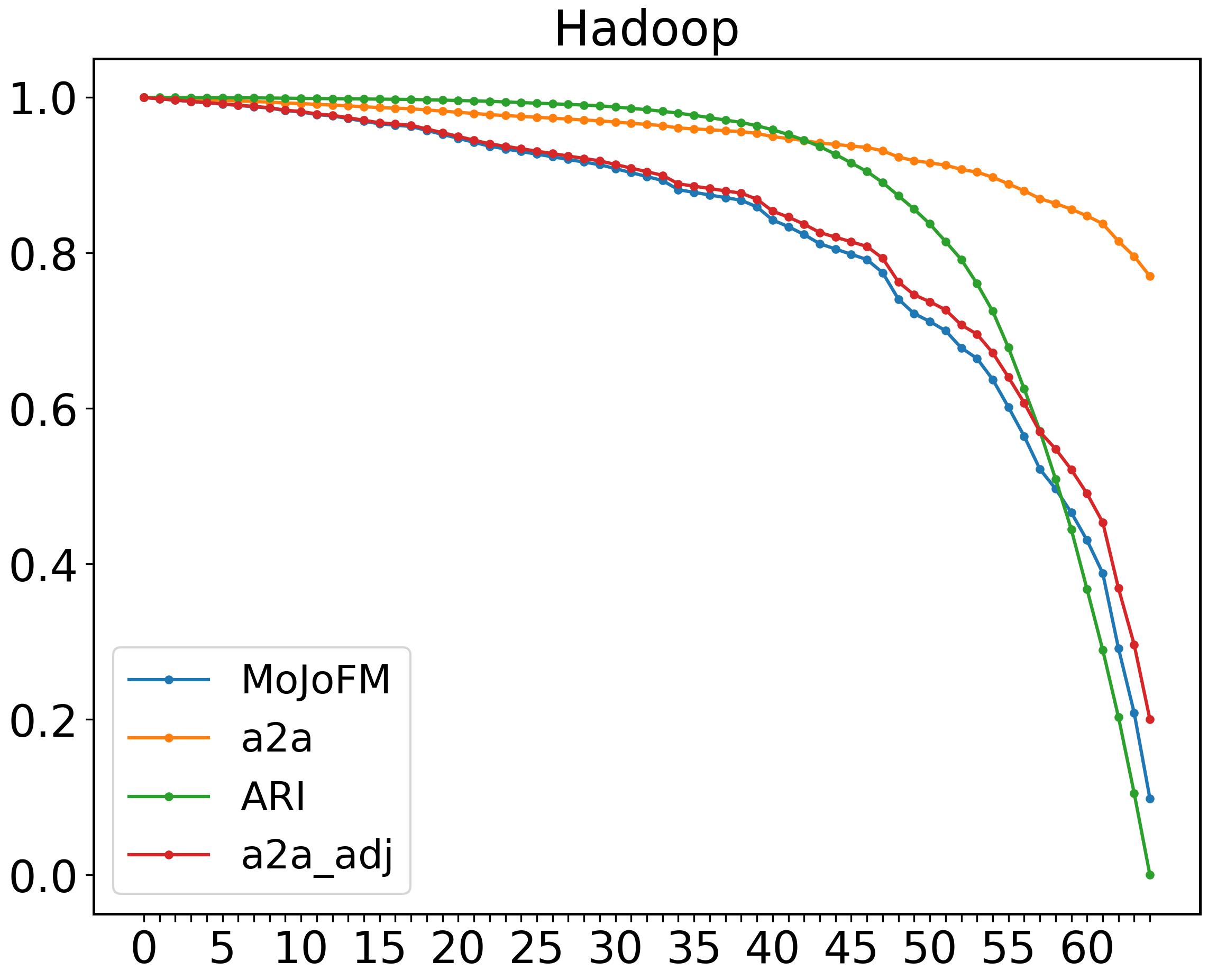}
        \label{subfig_metric_merge1}
    }
    \subfigure[$c2c_{cvg}$ w/ different thresholds]{
        \includegraphics[width=0.21\textwidth]{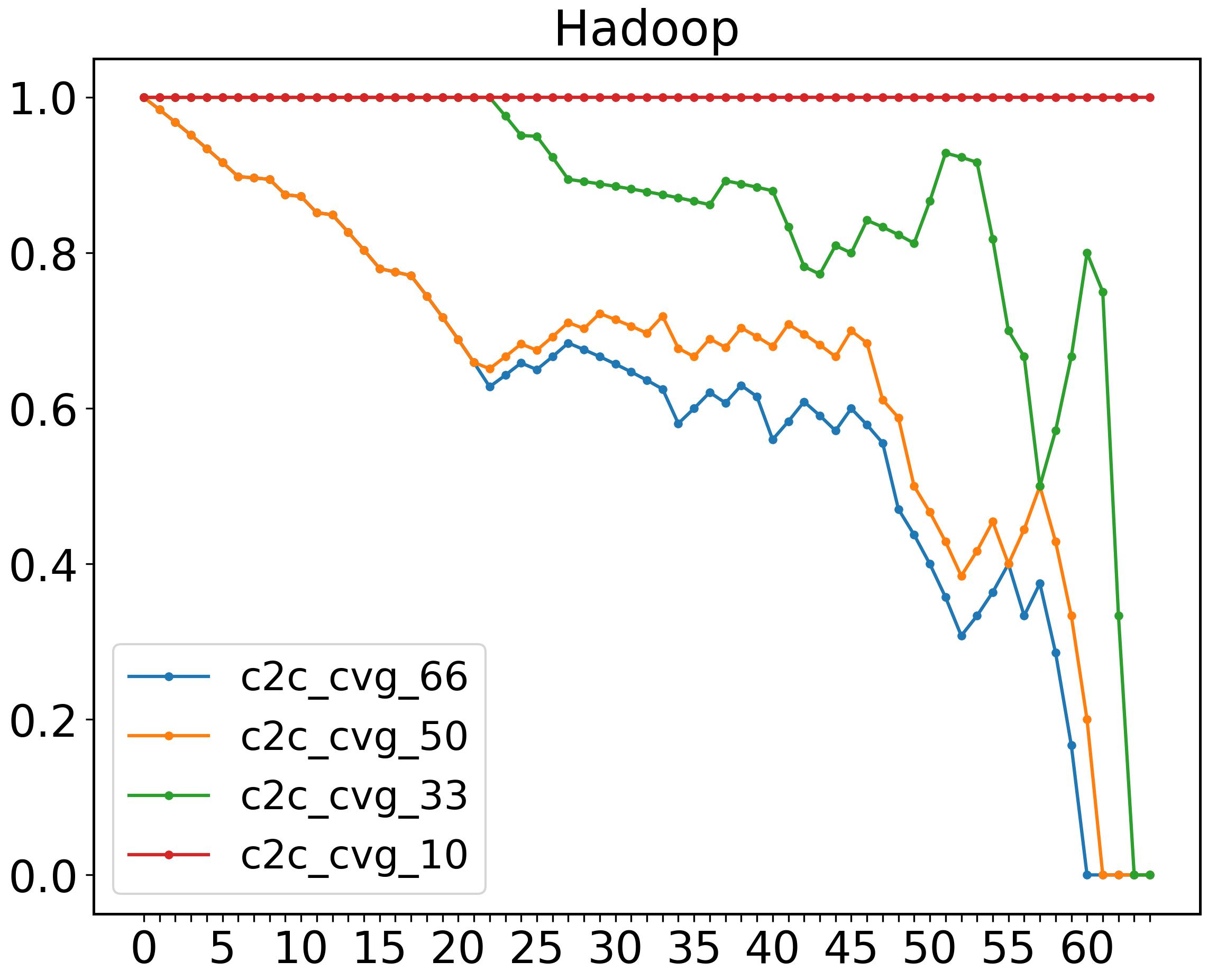}
        \label{subfig_metric_merge2}
    }
    \caption{Metrics Results of the Merge Experiment.}
    \label{fig_metric_merge}
\end{figure}

\begin{figure}[t]
    \centering    
    \subfigure[MoJoFM, a2a, ARI and $a2a_{adj}$]{
        \includegraphics[width=0.21\textwidth]{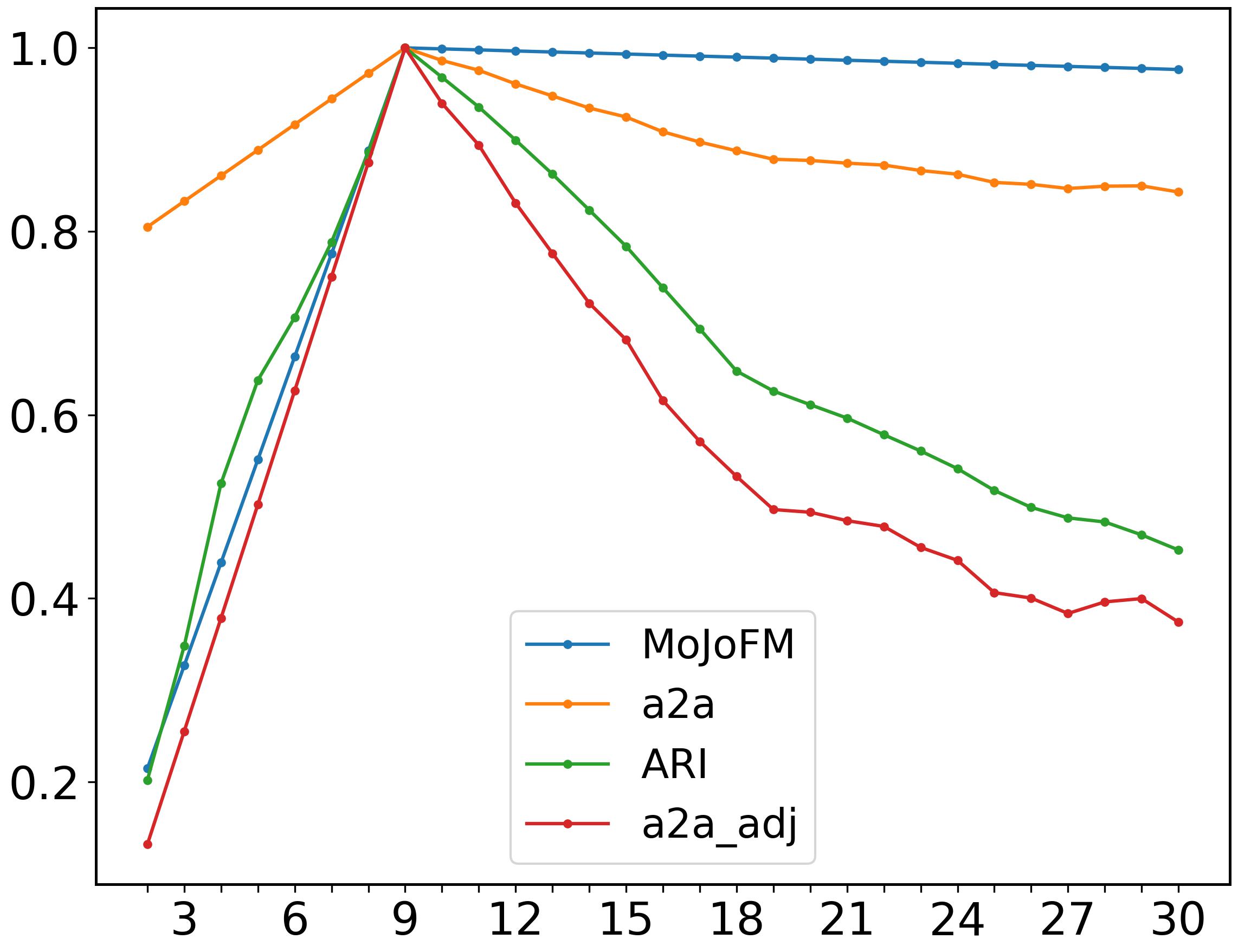}
        \label{subfig_metric_9test1}
    }
    \subfigure[$c2c_{cvg}$ w/ different thresholds]{
        \includegraphics[width=0.21\textwidth]{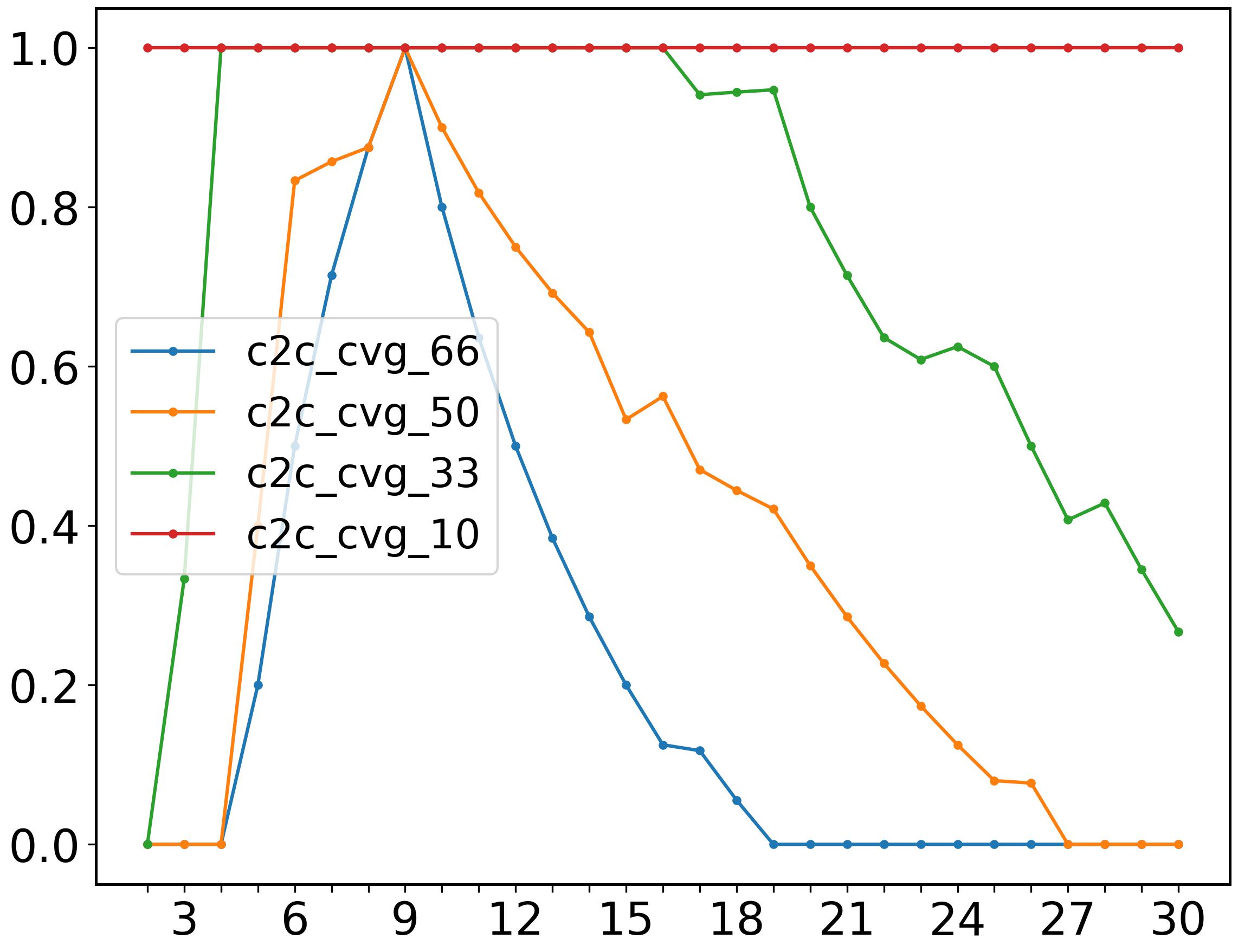}
        \label{subfig_metric_9test2}
    }
    \caption{Metrics Results of Nine Cluster Experiment.}
    \label{fig_metric_9test}
\end{figure}


{
\setlength{\tabcolsep}{1.85pt}
\begin{table*}[tb]
    \centering
    \caption{Similarities Between Recovered Architectures and Ground Truths. Top-2 Scores of each project are highlighted. All metric values are timed by 100. (Abbreviations: M-MoJoFM, A-a2a, C-\ccg, R-ARI, J-\aaj)} \footnotesize
    \label{tab_metrics_result}
    \begin{tabular}{l|ccccc|ccccc|ccccc|ccccc|ccccc|ccccc|ccccc|ccccc|ccccc}
        \hline
        Technique & \multicolumn{5}{c|}{ArchStudio} & \multicolumn{5}{c|}{Bash} & \multicolumn{5}{c|}{Chromium} & \multicolumn{5}{c|}{Hadoop} & \multicolumn{5}{c|}{ITK} & \multicolumn{5}{c|}{OODT} & \multicolumn{5}{c|}{HDC} & \multicolumn{5}{c|}{HDF} & \multicolumn{5}{c}{Libxml2}\\
        \cline{1-46}
        & {\scriptsize M} & {\scriptsize A} & {\scriptsize C} & {\scriptsize R} & {\scriptsize J} & {\scriptsize M} & {\scriptsize A} & {\scriptsize C} & {\scriptsize R} & {\scriptsize J} & {\scriptsize M} & {\scriptsize A} & {\scriptsize C} & {\scriptsize R} & {\scriptsize J} & {\scriptsize M} & {\scriptsize A} & {\scriptsize C} & {\scriptsize R} & {\scriptsize J} & {\scriptsize M} & {\scriptsize A} & {\scriptsize C} & {\scriptsize R} & {\scriptsize J} & {\scriptsize M} & {\scriptsize A} & {\scriptsize C} & {\scriptsize R} & {\scriptsize J} & {\scriptsize M} & {\scriptsize A} & {\scriptsize C} & {\scriptsize R} & {\scriptsize J} & {\scriptsize M} & {\scriptsize A} & {\scriptsize C} & {\scriptsize R} & {\scriptsize J} & {\scriptsize M} & {\scriptsize A} & {\scriptsize C} & {\scriptsize R} & {\scriptsize J}\\
        ACDC & \cellcolor{black!40}73 & \cellcolor{black!25}86 & \cellcolor{black!25}24 & 20 & \cellcolor{black!25}50 & \cellcolor{black!25}50 & 81 &  4 & 13 & 40 & 48 & 82 &  1 &  7 & 37 & 46 & 82 &  5 & 13 & 33 & 40 & 73 &  0 &  2 & 25 & \cellcolor{black!25}44 & \cellcolor{black!25}85 & 17 & \cellcolor{black!40}24 & \cellcolor{black!40}45 & 49 & 82 &  0 & 11 & 32 & 38 & 79 &  0 & 13 & 39 & 34 & 83 & \cellcolor{black!25}25 & 13 & 40\\
        ARC & 57 & 85 & 10 & \cellcolor{black!25}32 & 47 & 28 & 80 &  0 &  3 & 36 & 33 & 83 &  0 &  1 & 32 & 40 & 83 & 13 & 19 & 35 & \cellcolor{black!40}76 & 77 & \cellcolor{black!25} 0 &  0 & 32 & 30 & 82 & \cellcolor{black!25}17 & 12 & 38 & 37 & 81 &  0 &  8 & 37 & 36 & 80 &  0 & 12 & 38 & 29 & 83 &  6 &  6 & 32\\
        Bunch-NAHC & 39 & 82 &  6 & 14 & 34 & 39 & 83 &  0 &  9 & 26 & 57 & 73 &  0 &  0 & 20 & 34 & 82 &  0 & 12 & 31 & 40 & 80 &  0 &  1 & 15 & 14 & 77 &  0 &  6 & 29 & 55 & 86 &  0 & 24 & 37 & 42 & 84 &  0 & 11 & 35 & 22 & 81 &  0 &  9 & 36\\
        Bunch-SAHC & 62 & 85 &  7 & 22 & 41 & 45 & 83 &  4 & 12 & 31 & 27 & 74 &  0 &  0 & 12 & 34 & 81 &  0 & 15 & 33 & 54 & 78 &  0 &  4 & 14 & 14 & 77 &  0 &  6 & 29 & 65 & 88 &  0 & 34 & 48 & 40 & 86 &  0 & 17 & 38 & \cellcolor{black!25}38 & \cellcolor{black!25}86 & 11 & 20 & \cellcolor{black!25}47\\
        FCA & 60 & 82 &  9 & 10 & 43 & 41 & 80 &  2 & 13 & 40 & 58 & 76 &  0 &  3 & 29 & \cellcolor{black!25}49 & 82 &  7 & 11 & 38 & 44 & 73 &  0 &  1 & 25 & \cellcolor{black!40}50 & \cellcolor{black!40}86 & 14 & \cellcolor{black!25}23 & \cellcolor{black!25}44 & 53 & 78 &  0 & 10 & 36 & 32 & 78 &  0 & 15 & 42 & 37 & 84 & 10 &  8 & 37\\
        LIMBO & 27 & 80 &  0 & -0 & 17 & 26 & 80 &  0 &  0 & 22 & 31 & 78 &  0 &  0 & 11 & 15 & 80 &  0 &  1 & 18 & 50 & 78 &  0 &  0 &  6 & 12 & 79 &  0 &  0 & 22 & 25 & 80 &  0 &  1 & 30 & 20 & 81 &  0 & -0 & 31 & 22 & 84 &  3 &  2 & 43\\
        SADE & 42 & 83 &  0 & 11 & 41 & 45 & \cellcolor{black!25}86 & \cellcolor{black!40}17 & \cellcolor{black!25}21 & \cellcolor{black!25}42 & \cellcolor{black!25}58 & \cellcolor{black!25}87 & \cellcolor{black!40}24 & \cellcolor{black!25}26 & \cellcolor{black!25}48 & 41 & \cellcolor{black!25}83 & \cellcolor{black!25}22 & \cellcolor{black!25}23 & \cellcolor{black!25}44 & 60 & 81 &  0 & \cellcolor{black!25} 9 & 24 & 21 & 78 & \cellcolor{black!40}17 & 13 & 34 & \cellcolor{black!25}72 & \cellcolor{black!25}88 & \cellcolor{black!25} 8 & \cellcolor{black!25}41 & \cellcolor{black!25}53 & \cellcolor{black!25}49 & \cellcolor{black!25}88 & \cellcolor{black!40}22 & \cellcolor{black!25}30 & \cellcolor{black!25}47 & 34 & 84 & 17 & \cellcolor{black!25}22 & 44\\
        WCA-UE & 31 & 82 &  0 &  3 & 35 & 40 & 81 &  0 &  9 & 37 & 31 & 82 &  0 &  2 & 29 & 15 & 79 &  0 & -1 & 19 & 49 & 86 &  0 & -7 & 42 & 11 & 78 &  3 &  0 & 22 & 51 & 80 &  0 & 11 & 34 & 45 & 82 &  5 & 26 & 46 & 33 & 84 &  7 & 10 & 45\\
        WCA-UENM & 31 & 82 &  0 &  3 & 35 & 38 & 80 &  0 &  7 & 32 & 31 & 82 &  0 &  2 & 29 & 15 & 79 &  0 & -1 & 19 & 49 & \cellcolor{black!25}86 &  0 & -7 & \cellcolor{black!25}42 & 11 & 78 &  3 &  0 & 22 & 48 & 80 &  0 & 10 & 34 & 44 & 82 &  3 & 23 & 44 & 29 & 83 &  3 &  6 & 44\\
        \ourtool & \cellcolor{black!25}{\mj{67}} & \cellcolor{black!40}{\mj{88}} & \cellcolor{black!40}{\mj{36}} & \cellcolor{black!40}\mj{{42}} & \cellcolor{black!40}{\mj{55}} & \cellcolor{black!40}{\mj{70}} & \cellcolor{black!40}{\mj{87}} & \cellcolor{black!25}{ \mj{9}} & \cellcolor{black!40}{\mj{55}} & \cellcolor{black!40}{\mj{64}} & \cellcolor{black!40}{\mj{63}} & \cellcolor{black!40}{\mj{87}} & \cellcolor{black!25}{\mj{17}} & \cellcolor{black!40}{\mj{32}} & \cellcolor{black!40}{\mj{49}} & \cellcolor{black!40}{\mj{59}} & \cellcolor{black!40}{\mj{87}} & \cellcolor{black!40}{\mj{24}} & \cellcolor{black!40}{\mj{49}} & \cellcolor{black!40}{\mj{53}} & \cellcolor{black!25}{\mj{68}} & \cellcolor{black!40}{\mj{86}} & \cellcolor{black!40}{\mj{ 2}} & \cellcolor{black!40}{\mj{30}} & \cellcolor{black!40}{\mj{44}} & {\mj{29}} & {\mj{80}} & {\mj{14}} & {\mj{20}} & {\mj{41}} & \cellcolor{black!40}{\mj{90}} & \cellcolor{black!40}{\mj{90}} & \cellcolor{black!40}{\mj{33}} & \cellcolor{black!40}{\mj{59}} & \cellcolor{black!40}{\mj{62}} & \cellcolor{black!40}{\mj{56}} & \cellcolor{black!40}{\mj{88}} & \cellcolor{black!25}{\mj{13}} & \cellcolor{black!40}{\mj{35}} & \cellcolor{black!40}{\mj{50}} & \cellcolor{black!40}{\mj{49}} & \cellcolor{black!40}{\mj{89}} & \cellcolor{black!40}{\mj{33}} & \cellcolor{black!40}{\mj{29}} & \cellcolor{black!40}{\mj{55}}\\
        \hline
    \end{tabular}
\end{table*}
}

\noindent\textbf{Experiment 1:} 
To evaluate the ability of the metrics in measuring the architecture similarity, we design an experiment to simulate the situation where the target architecture becomes less and less similar to the ground truth. The expectation 
is that the metrics should give a lower score when the target architecture becomes dissimilar.
Specifically, we start by comparing the ground truth architecture of Hadoop with itself where all the metrics will give the highest score (i.e. 1.0). Then, we modify the architecture by merging its clusters and comparing it with the original ground truth again to get the metric scores. As we merge more clusters, the resulting architecture will be more different from the original one, to which a lower score should be given by the metrics. 

The results are shown in Fig~\ref{fig_metric_merge}, where the x-axis shows the number of clusters merged (i.e. the degree of dissimilarity) and the y-axis shows the metric scores.
Four metrics in Fig~\ref{subfig_metric_merge1} fit our expectation since their scores keep dropping during the merge. 
However, the score of a2a is not dropping as much as the rest. It complies with the conclusion in~\cite{lutellier2015comparing} that a2a has a relatively small variation range. It will always give a relatively good score for all kinds of architecture.
As for $c2c_{cvg}$ shown in Fig~\ref{subfig_metric_merge2}, the trend of score does not align well with our expectation. When the threshold is set to 0.1, the score is stucked at 1, which means $c2c_{cvg}$ could produce totally useless result if the threshold is set inproperly. Moreover, for the other thresholds, $c2c_{cvg}$ is overall dropping, but not stable enough: It fluctuates significantly during the merge process, which mismatches our intuitive expectation.



\noindent\textbf{Experiment 2:} In the second experiment, we aim to test the metric quality when the target architecture consists of different numbers of clusters. To achieve this, we generate a set of 9 well-separated clusters of data as the ground truth. Then, we use a k-means algorithm to cluster the data into 1-30 clusters as 30 different architectures. The architecture with 9 clusters is the same as the ground truth. Therefore, all the metric scores will be 1. If the number of clusters in the architecture becomes less or more than 9, the architecture will become different from the ground truth, which is expected to receive a lower score. 
The results are shown in Fig~\ref{fig_metric_9test}, where the x-axis shows the number of result clusters.


As expected, all the metrics give a score of 1 to the architecture with 9 clusters.
Similar to the last experiment, four metrics shown in Fig.~\ref{subfig_metric_9test1} are in line with our expectations. 
Specifically, MoJoFM gives all the architectures with more than 9 clusters a score close to 1, which suggests that it favours smaller clusters.
As for $c2c_{cvg}$ shown in Fig~\ref{subfig_metric_9test2}, though it can sometimes give a reasonable result, it still suffers from two limitations similar to the last experiment: 1) a proper threshold need to be selected, 2) the trend is not stable enough even if a proper threshold is set.

Based on the result of these two experiments, we do not find any substantial limitations of ARI and \aaj. For MoJoFM, it can also generate reasonable scores in most cases, though it favors partitions with a higher number of clusters. For a2a, the trend of score is satisfactory, but its dynamic range is limited. As for \ccg, its performance relies heavily on a proper threshold, while may still gives unreasonable results even if the threshold is appropriate.

\mybox{Answering to RQ1: 
ARI and $a2a_{adj}$ are the best metrics in the task of architecture similarity evaluation. MoJoFM, a2a, and $c2c_{cvg}$ have their limitation in particular scenarios.}


\subsection{Recovered Architecture Evaluation (RQ2)}
In this RQ, we evaluate \ourtool against baselines by comparing their recovered architectures against human-labeled ground truth.



\subsubsection{Metric Results}
We evaluate \ourtool and all the previous tools on the ground-truth projects with the setups detailed in Sec~\ref{sec:expset}. 
Although the outcome of RQ1 indicates that MoJoFM, a2a and \ccg have their limitations, all five metrics are included in our evaluation for the sake of completeness.
The threshold of \ccg is set to 0.66 since it gave the most reasonable result in RQ1. A total of 90 unique experiments are carried out to get all scores of each metric for each tool on each project.
The results of the metrics are shown in Table~\ref{tab_metrics_result}. Out of a total of 40 metrics on 8 projects, we achieved top-1 scores for 77.8\% (35/45) results, and top-2 scores for 88.9\% (40/45) results among all 10 algorithms. We summarized the averaged metric result in Table~\ref{tab_avg_metric}. The 'Best Base.' presents the best scores of the baseline techniques for each metric. 
\ourtool is \mj{36.1\%} more accurate than the previous best results in terms of the average of the leading percentage of all five metrics.

\begin{table}[t]
    \caption{Averaged Scores of Recovery Techniques.}\footnotesize
    \label{tab_avg_metric}
    \centering
    \begin{tabular}{l|ccccc}
        \hline
        Technique & MoJoFM & a2a & \ccg & ARI & \aaj\\ \hline\hline
        ACDC & 47 & 81 &  8 & 13 & 38\\ 
        ARC & 41 & 82 &  5 & 10 & 36\\ 
        Bunch-NAHC & 38 & 81 &  1 &  9 & 29\\ 
        Bunch-SAHC & 42 & 82 &  2 & 15 & 32\\ 
        FCA & 47 & 80 &  5 & 10 & 37\\ 
        LIMBO & 25 & 80 &  0 &  0 & 22\\ 
        SADE & 47 & 84 & 14 & 22 & 42\\ 
        WCA-UE & 34 & 82 &  2 &  6 & 35\\ 
        WCA-UENM & 33 & 81 &  1 &  5 & 34\\ \hline
        Best Base. & 47 & 84 & 14 & 22 & 42\\
        \ourtool & \mj{61} & \mj{87} & \mj{20} & \mj{39} & \mj{53}\\  \hline
        \ourtool v.s. Base. & \mj{+30.0\%} & \mj{+3.0\%} & \mj{+44.2\%} & \mj{+78.0\%} & \mj{+25.2\%}\\ \hline
    \end{tabular}    
\end{table}

\begin{table}[t]
    \caption{Result on OODT with Refined Granularity.}\footnotesize
    \label{tab_oodt_metric}
    \centering
    \begin{tabular}{l|ccccc}
        \hline
        Technique & MoJoFM & a2a & \ccg & ARI & \aaj\\ \hline\hline
        Pre. Best & 50 & 86 & 17 & 24 & 45\\ 
        \ourtool & \mj{56} & \mj{88} & \mj{19} & \mj{48} & \mj{55}\\ \hline 
        \ourtool v.s. Base. & \mj{+11.6\%} & \mj{+2.4\%} & \mj{+9.2\%} & \mj{+96.7\%} & \mj{+20.8\%}\\ \hline        
    \end{tabular}    
\end{table}

The only project we failed to achieve top 2 is OODT. 
The reason is that for OODT, the ground truth architecture has 216 clusters, while \ourtool only recovers the project into 28 clusters with the default $resolution$ parameter. Such a huge discrepancy leads to the low metric scores of \ourtool. As discussed in Sec~\ref{subsec_final_cluster}, our default resolution normally gives a result of 10 to 30 clusters which we think is suitable for humans to understand. However, there is no standard for the granularity of architecture since different granularity will be desired in different use cases. Our default granularity mismatches the ground-truth architecture of OODT, which is much fine-grained.
Therefore, we conducted an extensive experiment that increases the \textit{resolution} parameter by 20 times to make the granularity of \ourtool comparable to the ground truth of OODT. With the updated {resolution}, \ourtool clusters OODT into 180 clusters and the corresponding metrics are shown in Table~\ref{tab_oodt_metric}. With the updated \textit{resolution}, our result is much more accurate than the baseline techniques.

In contrast to our default coarse-grained granularity, other leading techniques like ACDC, FCA or SADE tend to decompose the system into much more clusters. This is the most common reason for the metrics that we fall behind. 
For example, on ArchStudio, ACDC results in 240 clusters while \ourtool generates only 25 clusters. As mentioned in RQ1, MoJoFM favors small clusters in nature. Consequently, ACDC beats us on ArchStudio for MoJoFM, while we achieved a higher score for the other four metrics.




\subsubsection{\mj{Time Consumptions}}
\mj{{To evaluate the scalability of \ourtool, the time consumption for the architecture recoveries is measured and shown in Table~\ref{tab_time}. The experiments are conducted on an Ubuntu server with dual 20-core Intel CPUs and 188 GB RAM. The results show that \ourtool can recover the architectures within a reasonable timeframe: small and medium size projects can be finished within a few minutes, while projects as large as Chromium, which has more than 10 million lines of codes, can be finished with about 80 minutes.
It can be found that the time consumption of \ourtool is about linearly proportional to the size of the project measured by the lines of code, which ensured our scalability on large projects.
\ourtool may take longer time than some alternative tools, since we utilized both dependency and textual information while other tools typically only use one dimension. The inclusion of these two dimensions inevitably increased the time cost since extracting them could be time-consuming. For example, among a total of 4499.3 seconds when recovering Chromium with \ourtool, 3943.3 seconds (87.6\%) were spent on the information extraction, while the other processes only take less than 10 minutes.}}

{
\setlength{\tabcolsep}{1.11pt}
\begin{table}[tb] 
    \caption{\mj{Time Consumption For Techniques (in seconds).}}\footnotesize
    \label{tab_time}
    \mj{\begin{tabular}{l|ccccccccc}
    \hline
    Technique& AS4 & Bash & Chromium & Hadoop & ITK    & OODT  & HDC  & HDF  & Libxml2 \\
    \hline
    ACDC     & 46.5        & 10.4  & 28,259.3  & 20.2   & 345.7  & 16.1  & 10.1  & 16.0   & 9.6     \\
    ARC      & 2,073.6      & 510.0 & 23,769.3  & 981.5 & 8,266.0 & 657.2 & 110.1 & 915.2 & 619.7   \\
    FCA      & 66.5        & 10.4  & 14,213.0  & 27.0   & 771.9  & 24.2  & 10.5  & 22.1   & 9.5     \\
    LIMBO    & 292.5       & 13.9  & 10,538.6  & 70.6   & 1976.0 & 33.6  & 11.5  & 27.7   & 10.1    \\
    SADE     & 37.4        & 10.6  & 3,813.0   & 19.9   & 272.0  & 16.4  & 10.7  & 15.9   & 10.2    \\
    WCA-UE   & 78.4        & 10.9  & 4,368.0   & 27.6   & 471.3  & 19.4  & 10.8  & 18.2   & 10.2    \\
    WCA-NM   & 82.9        & 11.0  & 4,386.3   & 28.8   & 473.0  & 19.3  & 10.8  & 18.3   & 10.1    \\
    Bunch-NAHC   & 50.3	&10.0	&160,969.6	&21.1	&5,765.8	&16.1	&9.9	&15.4	&9.4
    \\
    Bunch-SAHC   & 167.4	&10.3	&668,733.4	&33.9	&9,675.0	&20.2	&9.9	&18.2	&9.4    \\
    \ourtool & 92.2        & 74.5  & 4,499.3   & 110.1  & 1,195.6 & 84.3  & 22.0  & 44.1   & 45.9  \\
    \hline
    \end{tabular}}
\end{table}

}




\subsubsection{Case Study}
\begin{figure}[t]
    \centering    
    \subfigure[Reference Architecture~\cite{repo-dist-cam}]{
        \includegraphics[width=0.235\textwidth]{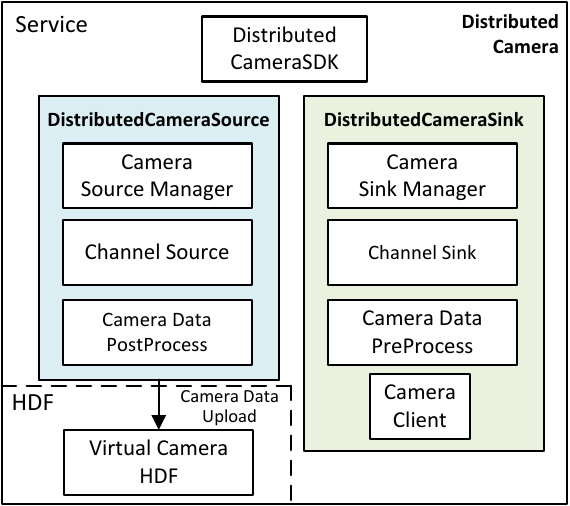}
        \label{fig_arch_dist_cam}
    }
    \subfigure[Recovered v.s. Ground Truth]{
        \fbox{\includegraphics[width=0.215\textwidth]{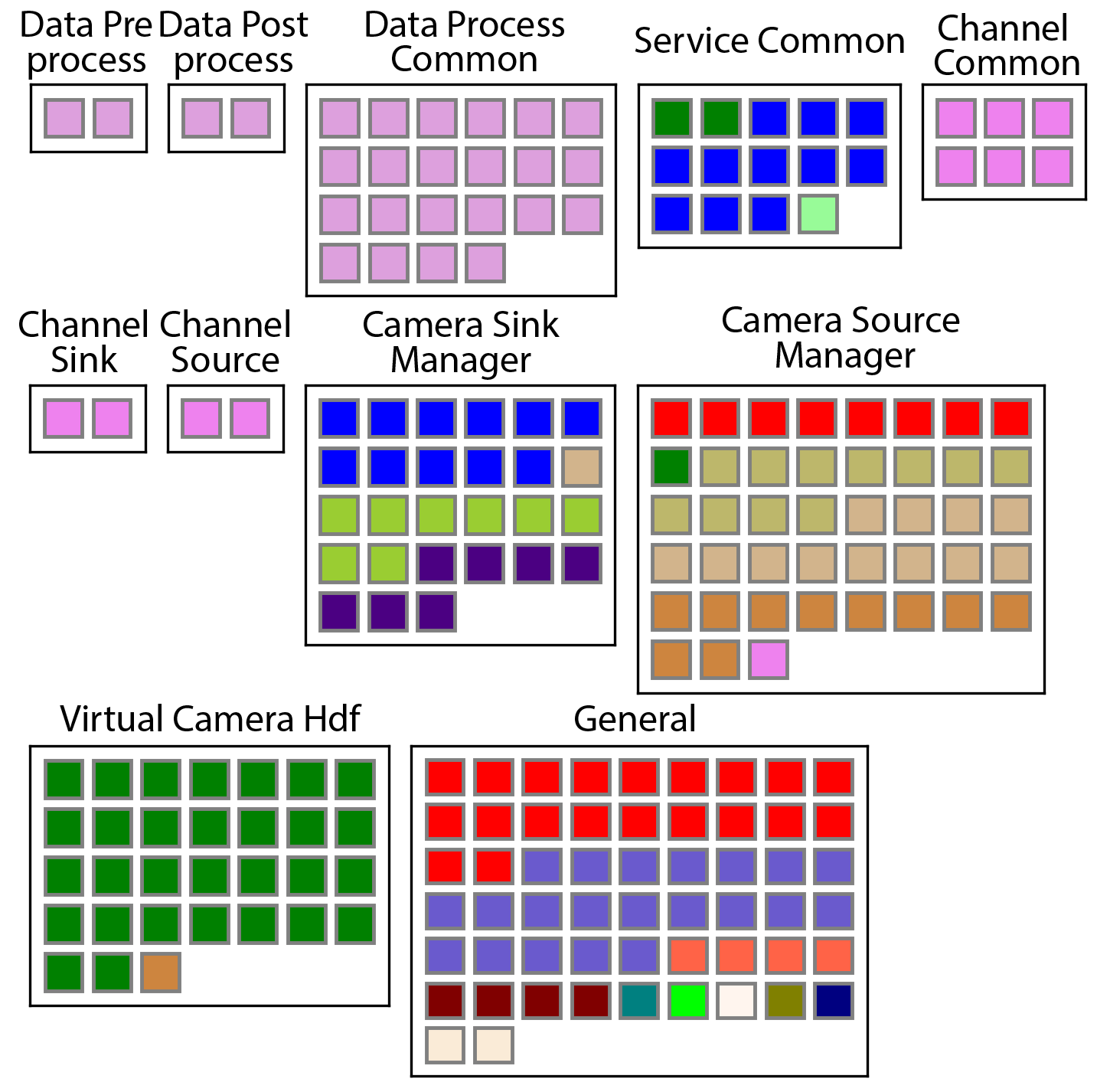}}
        \label{fig_res_dist_cam}
    }
    \caption{Reference Architecture of Distributed Camera, and Comparison between the Recovered Architecture and the Ground Truth.}
    \label{fig_dist_cam}
\end{figure}

To examine the practicalness of \ourtool, we perform a case study, which compares the architecture recovery results of \ourtool against the ground truth labeled by the developers of a real-world project Distributed Camera~\cite{repo-dist-cam}.
Fig~\ref{fig_arch_dist_cam} shows its architecture diagram from its source repository. The diagram reveals that it has two primary components: the camera source and the camera sink. Both the source and sink are comprised of three components: the manager, channel and data-processing.
However, while labeling the ground truth architecture based on the diagram, we find it hard to split the implementation of source and sink. For example, a total of 26 source files are related to data-processing, but 22 of them are common implementations shared by both the source and sink sides. 
Following the architecture diagram, we labeled the data-processing files into 3 modules: source, sink and common, though there are only 4 files in the source and the sink part. The other modules with similar issues like channel-related codes are also marked in this way. Last, we labeled and put the files that we cannot decide on their module with the diagram into a general module.

We then recover its architecture with \ourtool. The result is shown in Fig~\ref{fig_res_dist_cam} with distribution map~\cite{ducasse2006distribution}. 
Each colored small square represents a file. The files in the same box mean that they belong to the same human-labeled ground truth module, and the color-coding represents the result of \ourtool. As shown in the figure, \ourtool successfully identifies the data-processing related module (colored light pink) and channel module (colored pink), i.e., \ourtool clusters file with similar functionalities like data-processing together rather than splitting them by source and sink.
Although this result does not align with the reference architecture, we believe that the result of \ourtool is even more reasonable. As previously stated, 22 out of 26 files in the data-processing module are shared by both source and sink. From an implementation perspective, it is unfair to split the module just because the minority four files can be well separated.


\subsubsection{\mj{Extensive Validation.}} 

    To further validate the generalizability of \ourtool, an extensive experiment is carried out on 900 GitHub projects as detailed in Sec~\ref{subsubsec_data_collection}.
    To quantitatively compare the result of \ourtool with baselines, a ground truth architecture is required. However, labeling a reliable ground truth architecture from scratch requires extremely high manual efforts and extensive communication with the corresponding developers, which is time-consuming. For example, labeling Chromium took 2 years from the previous researcher~\cite{lutellier2015comparing}. 
    Given our aim is to generate more ground truth data to test the generalizability of \ourtool, this accurate but time-consuming method is not feasible. Therefore, we adopted an alternative labeling method. For each repository, if it contains a reference architecture diagram, we labeled the architecture accordingly. While this alternative method may not be as accurate as traditional labeling, it should sufficiently represent the developers' architectural perception.
    
    Using this method, we labeled a total of 23 projects. We compared \ourtool with baselines based on these labeled architectures. Metric results indicate \ourtool's accuracy exceeds baselines by 35.2\%. All data related to this extensive experiment, including project list, labeled architecture, metric results and recovered architecture for all 900 projects are available on our website~\cite{oursite} for public verification.

\mybox{Answering RQ2: \ourtool has the best architecture recovery results compared to 9 related works with a \mj{36.1\%} higher in accuracy on average. \mj{An extensive experiment verified \ourtool's generalizability.} In the case study, it produces a more reasonable architecture than the human-label one.}


\subsection{Ablation Study (RQ3)}
In this RQ,  we aim to evaluate the impact of two factors on the accuracy of \ourtool: 1) fusing various types of information, and 2) using fine-grained dependency weighting. Regarding the information fusion, we will investigate the impact of excluding textual information or folder structure, while dependency cannot be totally excluded since our weight assignment algorithm is based on it. 
As for the weighting of dependencies, we will evaluate the impact of removing the weights for dependency types and entity importance.

\begin{table}[tb]
    \caption{Accuracy with Incomplete Information/Weight.}\scriptsize
    \label{tab_metric_incomp}
    \begin{tabular}{p{1.5cm}|p{1.1cm} c c c c c}
        \hline
        Variant&& MoJoFM & a2a & \ccg & ARI & \aaj \\  
        \hline \hline
        \multirow{3}*{\ourtool}&avg& \mj{61}& \mj{87}& \mj{20}& \mj{39}& \mj{53}\\
        &v.s. \ourtool& /& /& /& /& /\\
        &v.s. best base.& \mj{+30.0\%}& \mj{+3.0\%}& \mj{+44.2\%}& \mj{+78.0\%}& \mj{+25.2\%}\\ \hline \hline  
        \multirow{3}*{w/o Folder}&avg& \mj{51}& \mj{85}& \mj{15}& \mj{27}& \mj{43}\\
        &v.s. \ourtool& \mj{-16.0\%}& \mj{-1.9\%}& \mj{-25.5\%}& \mj{-30.6\%}& \mj{-17.9\%}\\
        &v.s. best base.& \mj{+9.2\%}& \mj{+1.1\%}& \mj{+7.4\%}& \mj{+23.5\%}& \mj{+2.9\%}\\ \hline      
        \multirow{3}*{w/o Text}&avg& \mj{52}& \mj{85}& \mj{13}& \mj{29}& \mj{47}\\
        &v.s. \ourtool& \mj{-15.4\%}& \mj{-2.5\%}& \mj{-35.3\%}& \mj{-26.0\%}& \mj{-10.8\%}\\
        &v.s. best base.& \mj{+10.0\%}& \mj{+0.4\%}& \mj{-6.7\%}& \mj{+31.7\%}& \mj{+11.6\%}\\ \hline    
        \multirow{3}*{Dep. Only}&avg& \mj{49}& \mj{83}& \mj{ 7}& \mj{22}& \mj{43}\\
        &v.s. \ourtool& \mj{-19.4\%}& \mj{-4.3\%}& \mj{-67.2\%}& \mj{-43.4\%}& \mj{-19.2\%}\\
        &v.s. best base.& \mj{+4.8\%}& \mj{-1.4\%}& \mj{-52.6\%}& \mj{+0.7\%}& \mj{+1.2\%}\\ \hline \hline  
        \multirow{3}*{w/o Entity Impt.}&avg& \mj{56}& \mj{86}& \mj{19}& \mj{32}& \mj{47}\\
        &v.s. \ourtool& \mj{-8.6\%}& \mj{-1.2\%}& \mj{-8.6\%}& \mj{-19.1\%}& \mj{-11.4\%}\\
        &v.s. best base.& \mj{+18.8\%}& \mj{+1.8\%}& \mj{+31.8\%}& \mj{+44.0\%}& \mj{+11.0\%}\\ \hline   
        \multirow{3}*{w/o Dep. Type}&avg& \mj{60}& \mj{85}& \mj{14}& \mj{31}& \mj{46}\\
        &v.s. \ourtool& \mj{-2.5\%}& \mj{-2.2\%}& \mj{-31.8\%}& \mj{-21.6\%}& \mj{-12.6\%}\\
        &v.s. best base.& \mj{+26.7\%}& \mj{+0.8\%}& \mj{-1.7\%}& \mj{+39.5\%}& \mj{+9.4\%}\\ \hline     
        \multirow{3}*{w/o Dep. Weight}&avg& \mj{55}& \mj{85}& \mj{12}& \mj{28}& \mj{43}\\
        &v.s. \ourtool& \mj{-10.4\%}& \mj{-2.2\%}& \mj{-41.0\%}& \mj{-27.1\%}& \mj{-17.5\%}\\
        &v.s. best base.& \mj{+16.5\%}& \mj{+0.8\%}& \mj{-14.9\%}& \mj{+29.8\%}& \mj{+3.2\%}\\ \hline
    \end{tabular}
\end{table}

Table~\ref{tab_metric_incomp} summarizes the average decrease in accuracy across 9 projects after removing information or weighing, and the \mj{comparison} between incomplete \ourtool and the best of the baseline techniques. As shown in the table, \ourtool's accuracy always decreases significantly after removing each type of information or weight.
The removal of folder structure has the greatest negative impact on most metrics. The scores of different metrics declined by about 16\% to 30\% (except for a2a, whose variation range is too small as discussed in RQ1). This  aligns with our expectations since some of the human-labeled ground truth architectures are strongly related to their folder structures~\cite{garcia2013obtaining}.
Removing textual information also reduces the scores by about 10\% $\sim$ 30\%, showing that the textual similarities can strongly indicate the architectural modules. 
When both textual information and folder structure are removed, our metric scores become comparable to the best baseline scores. This suggests that fusing multiple information sources is necessary for \ourtool to outperform the baselines.


As for dependency weighing, removing weight by entity importance or dependency type leads to a decrease in accuracy of about 3\% to 20\%. Removing both weighting results in a drop of about 11\% to 26\% in accuracy, which is comparable to the decrease by removing textual information or folder structure. This result confirms the importance of using fine-grained dependency weighting.





\mybox{Answering RQ3: Both information fusion and fine-grained dependency weighting are critical to \ourtool. Removing either  dependency weighting or one type of information will result in an accuracy drop of approximately 10\% to 30\% for different metrics. Removing two types of information will result in a low accuracy that comparable to the best baseline techniques.}

\balance
\section{Threats to Validity}\label{sec_threat}

We discuss the threats to the validity of our results to comprehend the strength and limitations of \ourtool.
The first potential threat to validity relates to the representativeness and generalizability of our evaluation on \ourtool. The ground-truth architectures used in the evaluation only span 9 projects, which might not seem sufficient for robustly inferring generalizability. To alleviate this concern, we have gathered all publicly available ground-truth architectures from previous studies, and labeled several additional ground-truths with our industrial partners.
Moreover, an extensive evaluation is carried out on 900 Github projects to enhance the generalizability. 

The second threat pertains to the parameter selection within \ourtool. Assuring the optimal parameter setup for \ourtool
in the context of architecture recovery, is difficult. We have strived to make our design more reasonable by following the original algorithm designer's suggestions. Nevertheless, it is still uncertain if our parameter selections are the best.

The final threat related to the accuracy of dependency extraction. Even though DEPENDS is one of the most accurate open-source dependency extractors~\cite{Lefever2021On}, the dependency extracted by DEPENDS could be inaccurate due to the complex nature of programming languages and the limitation of static analysis.

\section{Related Work}\label{sec_related}

The architecture recovery techniques are to automatically recover software architectures from their implementations.
Many of these techniques use structural dependencies as their information source.
Bunch~\cite{mancoridis1999bunch} clusters source files of software systems by maximizing the modularity quality (MQ) of the dependency graph. 
WCA~\cite{maqbool2004weighted} is a hierarchical clustering technique that commonly-used in the scenario of software clustering. LIMBO~\cite{andritsos2004limbo} is another commonly used technique employing information theory concepts for software clustering.
DAGC~\cite{parsa2005new} is a technique that similar to Bunch but optimized its search space. 
MCA and ECA~\cite{praditwong2010software} introduced multi-objective optimization into software clustering.
Cooperative clustering technique (CCT)~\cite{naseem2013cooperative} is a consensus-based clustering technique that is utilized in the software clustering scenario. 
Mohammadi et al.~\cite{mohammadi_new_2019} used available knowledge in the dependency graph to create a neighborhood tree to drive the clustering.
In addition to these techniques that use static dependencies, Xiao et al.~\cite{xiao2005software} shows that dynamic dependencies have some merits.

Apart from dependency-based techniques, most studies use textual information for architecture recovery. ARC~\cite{garcia2011enhancing} recovers the architecture using concerns. ZBR~\cite{corazza2011investigating} is a recovery technique based on natural language semantics, and it partitions textual information into different zones to form clusters. 
Risi et al.~\cite{risi2012using} uses LSI to extracts textual information then cluster the system with K-Means.
Kargar et al.~\cite{kargar_semantic-based_2017} constructed a semantic dependency graph to replace the dependency graph to take the advantage that the semantic information is independent of programming languages.
They further employed nominal information (i.e. the name of artifacts) to improve their previous work in~\cite{kargar_multi-programming_2019}.
EVOL~\cite{yang_enhancing_2022} is a semantic-based technique that considers semantic outliers filtration and label propagation to increase their accuracy.
The textual-information-based techniques may outperform dependency-based ones in specific scenarios as shown in study~\cite{garcia2013comparative}.

There are also some techniques that use hybrid information for clustering.
Mkaouer et al.~\cite{mkaouer2015many} proposed a many-objective search-based approach using NSGA-III. 
Chhabra et at.~\cite{chhabra2017improving} extracted combined features with 24 different coupling schemes to explore how to combine the dependencies
and textual to produce the optimal clustering result.
SADE~\cite{cho_software_2019} clusters software systems by combining
the dependency information with the semantic similarity. They use the semantic similarity between modules as the weight of call graph, and cluster the graph using Louvain clustering algorithm.
Jalali et al.~\cite{sadat2019multi}  unified dependency information and semantic information to introduce a new multi-objective fitness function, and consider the clustering as a multi-objective search problem. 
However, current research on clustering with hybrid information mainly focus on only the dependency and textual information, while neglects many other useful information in the source codes.




\section{Conclusion}\label{sec_conclusion}  

In summary, we propose \ourtool, an architecture recovery technique
dedicates to comprehensively understanding the architecture of software systems by fusing three aspects of information from software systems. We evaluate \ourtool and 9 previous architecture recovery techniques on 9 software systems with ground truth architectures. Three of the most commonly-used metrics and two new metrics are utilized for evaluation. The experimental result shows that \ourtool outperforms the best baseline techniques by \mj{36.1\%} on average. A case study on a real-world software system shows that \ourtool can effectively identify modules with similar functionalities.
\mj{For future work, \ourtool can be further enhanced by incorporating more domain knowledge on architecture such as commonly-adopted architectural patterns.}
\section{Data Availability}
The data, demo and detailed instructions are available at \url{https://github.com/anonymous2f4a9d/SARIF_FSE23}.

\section*{Acknowledgementa}
This research / project is supported by the National Research Foundation, Singapore, and the Cyber Security Agency under its National Cybersecurity R\&D Programme (NCRP25-P04-TAICeN). Any opinions, findings and conclusions or recommendations expressed in this material are those of the author(s) and do not reflect the views of National Research Foundation, Singapore and Cyber Security Agency of Singapore.



\bibliographystyle{ACM-Reference-Format} 
\bibliography{ref-arch-rec}

\end{document}